# Evaluating Physician-AI Interaction for Cancer Management: Paving the Path toward Precision Oncology

Zeshan Hussain, MD, PhD*

Computer Science and Artificial Intelligence Laboratory, Massachusetts Institute of Technology

Barbara Lam, MD*

Division of Hematology and Oncology, University of Washington School of Medicine

Fernando A. Acosta-Perez

Department of Industrial and Systems Engineering, University of Wisconsin-Madison

Irbaz Bin Riaz, MD

Division of Hematology and Oncology, Department of Medicine, Mayo Clinic, Arizona

Maia Jacobs, PhD

Department of Computer Science, McCormick School of Engineering, Northwestern University

Andrew J. Yee, MD

Division of Hematology and Oncology, Department of Medicine, Massachusetts General Hospital

David Sontag, PhD

Computer Science and Artificial Intelligence Laboratory, Massachusetts Institute of Technology

As machine learning (ML)-based decision support tools proliferate in clinical practice, understanding how clinicians integrate personalized ML predictions alongside randomized controlled trial (RCT) evidence is critical. We designed a web-based clinical decision support system (CDSS) presenting survival and adverse event data from a simulated RCT and ML model across 12 synthetic multiple myeloma scenarios. In a within-subjects study with 32 physicians, we evaluated how clinicians synthesize competing evidence sources to make treatment decisions. When ML and RCT outputs were concordant, physicians reported greater confidence than with RCT data alone. When results were discordant, most physicians shifted toward the ML-supported treatment, often before reviewing any information about model training or validation, suggesting a tendency toward automation bias rather than algorithm avoidance. Despite reporting higher perceived reliability after viewing model quality disclosures, physicians were largely unable to describe the validation procedures they had reviewed. Taken together, these findings reveal that clinicians may over-rely on ML recommendations even when equipped with tools designed to support critical appraisal. We discuss implications for CDSS design, clinician training, and the institutional safeguards needed before ML-based systems are deployed in high-stakes oncology settings.

CCS CONCEPTS • Machine learning • Human-centered computing • Health care information systems

**Additional Keywords and Phrases:** Randomized controlled trials (RCTs), Clinical decision support tools, Multiple myeloma management

---

*equal contribution

## 1 INTRODUCTION

Cancer treatment has changed dramatically over the past decade. Advances in molecular sequencing and innovative therapies such as immune checkpoint inhibitors, bispecific antibodies, and CAR T-cells have significantly transformed the treatment landscape, namely by redefining standard of care across numerous tumor types and achieving durable responses in previously refractory hematologic malignancies [1–3]. Historically, the field was driven by landmark clinical trials [4–11]. Today, the highly personalized nature of each patient's disease and the expanding array of treatment options have made it increasingly difficult to capture all patient scenarios in clinical trials.

Multiple myeloma (MM) is often highlighted as an area where the rapid development of new therapies has outpaced our ability to conduct clinical trials [12, 13]. In the last ten years, the FDA has approved over a dozen new agents for MM [14]. However, registry studies show that about 40% of patients with MM did not meet the inclusion criteria for the phase 3 trials that supported those treatments [13]. Although randomized controlled trials (RCTs) are considered the gold standard in medical evidence [15], the gap between clinical trial and real-world practice will likely widen in MM, where clinicians use complex, multi-drug regimens and patients often require multiple lines of therapy [12]. Furthermore, this gap will be made even more evident on those patients, e.g. people with poor performance status, significant renal impairment, uncontrolled comorbidities, or advanced age, that are excluded from pivotal phase 3 clinical trials. For these patients, clinicians must extrapolate from trial data, drawing on clinical experience and indirect evidence.

This extrapolation gap is precisely where clinical decision support systems (CDSSs) provide the greatest value, offering a path to precision oncology by presenting personalized predictions derived from broader, real-world datasets. Modern CDSSs leverage machine learning (ML) models trained on large datasets to generate personalized treatment recommendations. While several CDSSs for cancer have been proposed and published in the literature [16-20], few make it to the bedside. Key barriers include the lack of prospective clinical validation, poor integration with existing electronic health record (EHR) workflows, clinician skepticism about "black-box" models, and regulatory uncertainty [21]. Surveys of clinicians indicate that they are generally optimistic about the role of AI in diagnostics and prognostication but express significant concerns about model validation and the potential for bias [22-24]. Studies have shown that clinicians may anchor on ML recommendations even when inaccurate [25], that perceived trustworthiness varies by clinical context and training level (though still may not impact reliance) [26], and that prescriptive recommendations bias decisions more than descriptive presentation of predicted outcomes [27]. However, little work has been done to understand how clinicians interact with ML-based systems in conjunction with current standard-of-care evidence from RCTs. Understanding this interplay is essential to support the implementation of ML-based CDSSs in real-world clinical settings.

In oncology, clinicians need to synthesize all available evidence to maximize patient survival and quality of life while minimizing adverse events. As ML systems become more prevalent, clinicians will increasingly need to reconcile findings from RCTs and ML models. If an ML model produces similar outcomes to an RCT, a clinician may feel more confident in their decision. Inevitably, however, an ML model will produce outcomes that contradict RCT data. To evaluate how clinicians navigate these complex scenarios, we designed a web-based CDSS that displays survival curves and adverse event data from a simulated RCT and a simulated ML model. We conducted a within-subjects experiment with 32 physicians who evaluated 12 synthetic patient scenarios, each presenting tiered information that progressively disclosed RCT outcomes, ML model predictions, and model quality metadata. Using a mixed-methods approach combining quantitative analysis of treatment choices and confidence ratings with qualitative thematic analysis of free-text responses and semi-structured interviews, we examined how clinicians integrate available evidence to make treatment decisions. In this study, we present the following contributions:

1) Design of a controlled experimental platform for studying physician-AI interaction in oncology, using a tiered information disclosure paradigm within a web-based CDSS
2) Empirical evidence on how concordance and discordance between RCT and ML outputs affect clinician confidence and treatment selection across systematically varied clinical scenarios.
3) Qualitative insights into clinicians' reasoning processes when integrating AI-generated evidence, including the identification of factors that promote trust and potential over-reliance vs skepticism.

4) Evidence that clinicians respond to model quality disclosures behaviorally but cannot articulate their content, revealing a gap between trust calibration and comprehension

### 1.1 Background: Multiple Myeloma

Multiple myeloma is a clonal plasma cell neoplasm characterized by the overproduction of monoclonal immunoglobulin, leading to end-organ damage including bone disease, renal insufficiency, anemia, and hypercalcemia. It is the second most common hematologic malignancy, with an estimated 35,000 new cases diagnosed annually within the United States [6]. Disease staging relies on the International Staging System (ISS) and its revised version (R-ISS), which incorporate serum beta-2 microglobulin, albumin, lactate dehydrogenase (LDH), and cytogenetic abnormalities to stratify patients by risk [6]. Standard-of-care treatment for newly diagnosed MM typically involves induction therapy with a multi-drug regimen combining a proteasome inhibitor (e.g., bortezomib), an immunomodulatory drug (e.g., lenalidomide), and a corticosteroid (e.g., dexamethasone), often followed by autologous stem cell transplantation for eligible patients and maintenance therapy [6,8,11]. As mentioned previously, the rapid approval of new agents, e.g. monoclonal antibodies, bispecific antibodies, and CAR T-cell therapies, has made treatment selection increasingly complex [14]. Clinicians must weigh disease biology (e.g., cytogenetic risk), patient fitness (e.g., performance status, comorbidities), treatment toxicity, and patient preferences when selecting among available regimens.

### 1.2 Background: The Clinical Information Ecosystem

Oncologists draw on a diverse set of information resources when making treatment decisions, including their clinical experience, national guidelines (e.g., NCCN Clinical Practice Guidelines [6]), published literature accessed through PubMed or UpToDate, institutional tumor board discussions with multidisciplinary colleagues, electronic health record (EHR) data on the patient's history and comorbidities, and direct conversations with patients about their goals of care. ML-based CDSSs are a new addition to this ecosystem. They offer individualized, data-driven predictions that complement population-level evidence from RCTs. Our study examines how clinicians integrate this new information source alongside RCT evidence in a simulated setting, recognizing that in clinical practice, CDSS outputs would be one input among many.

## 2 RELATED WORK

### 2.1 Clinical Decision Support Systems in Oncology

CDSSs have been developed for a range of oncology applications, from diagnostic support to treatment selection and prognostication [16-20]. In MM specifically, ML models have been proposed for predicting treatment response, survival, and adverse events [16, 43]. Kawamoto et al. [45] identified four features critical to CDSS effectiveness: automatic integration into workflow, delivery at the point of decision-making, actionable recommendations, and a computer-based platform. However, these principles were established for rule-based systems and do not address challenges specific to ML-driven tools.

Despite growing technical capabilities, adoption of AI-driven CDSSs in clinical practice remains limited [21]. Key barriers include insufficient prospective validation, poor integration with clinical workflows, and difficulty ensuring generalizability across patient populations [21]. Even technically accurate systems can fail in deployment. For example, Beede et al. [51] found that a deep learning system for diabetic retinopathy screening was undermined by infrastructure constraints and workflow disruptions in real clinics. However, this literature has largely focused on either the technical performance of CDSSs or the logistics of their deployment. Less attention has been paid to how clinicians reason about and integrate CDSS outputs into treatment decisions, particularly when those outputs must be weighed against existing evidence.

### 2.2 Trust, Over-reliance, and Explainability in Human-AI Interaction

A central challenge in human-AI collaboration is achieving appropriate reliance, defined as calibrating trust to match a system's actual capabilities [50]. Both over-trust and under-trust represent calibration failures. Automation bias, the tendency to defer to automated outputs uncritically, persists even among expert users and is not reliably eliminated by training [47]. Conversely, algorithm aversion leads people to disproportionately reject algorithmic predictions after observing errors, even when the algorithm outperforms human judgment [48]. Both failure modes have been documented, but primarily in non-clinical or single-source settings. Our study provides quantitative evidence of automation bias and qualitative evidence consistent with algorithm aversion operating among physicians making treatment decisions with two competing evidence sources.

In clinical AI specifically, how recommendations are framed affects clinician behavior. Adam et al. [27] showed that prescriptive advice produces greater decision bias than descriptive evidence, motivating our choice to present ML outputs as outcome predictions rather than treatment recommendations. Jacobs et al. [25] found evidence of anchoring on ML recommendations, and Gaube et al. [26] demonstrated that susceptibility to AI influence varies with framing. Bucinca et al. [42] proposed cognitive forcing functions as a mechanism to reduce over-reliance by requiring analytical engagement before viewing AI output. These studies examined clinician responses to AI as the primary or sole decision aid, though do not consider settings with concurrent, established evidence sources.

Research on explainability suggests that model explanations do not uniformly improve decision quality. Increased transparency can make errors harder to detect [37] and explanations may obscure biases [38]. Bansal et al. [52] found that AI explanations increased acceptance of recommendations regardless of their correctness. Cai et al. [49] found that clinicians encountering a diagnostic AI wanted global model-level information (e.g. training data composition, known limitations, design objectives) before engaging with case-specific predictions. While Cai et al. identified these onboarding needs qualitatively, our study tests their implications experimentally. Specifically, our tiered design progressively discloses the specific types of model-level information clinicians may find useful, and measures whether this disclosure changes treatment decisions and confidence.

### 2.3 Reconciling RCT and ML evidence

The studies reviewed above have examined clinician interactions with AI systems in isolation without a competing evidence source. In practice, however, ML-based CDSSs will not replace RCTs but supplement them, and the two may produce concordant or discordant recommendations. How clinicians reconcile these potentially conflicting inputs is an open question with direct implications for patient safety and CDSS design. Our study addresses this gap by systematically varying the concordance between RCT and ML outputs across different clinical scenarios, examining how agreement and disagreement between evidence sources affect treatment selection confidence and perceived model reliability.

Our quantitative approach follows an established tradition of within-subjects experimental designs in HCI research on human-AI collaboration, in which each participant serves as their own control to isolate the effect of AI-presented information on decision-making [25, 42, 52]. For the qualitative component, we applied Braun and Clarke's thematic analysis [28], a method that has been used to characterize clinician reasoning strategies and concerns when encountering AI systems in clinical contexts [49].

## 3 METHODS

### 3.1 Study Design

We employed a within-subjects design in which all participants evaluated 12 patient scenarios (A-L) in a randomized order. Each participant interacted with a web-based CDSS (Figure 1) that displayed survival curves (presented as progression-free survival) and adverse event data (presented as symptom frequencies) from a simulated RCT and a simulated ML model. These two outcomes, survival and adverse events, were chosen because they represent the primary axes along which oncologists evaluate treatment options, and are the standard endpoints reported in both RCTs [8, 11] and prognostic ML models in oncology [16-20]. At each tier of information, participants selected a treatment ("treatment red" or "treatment blue"), rated their confidence in their treatment choice, and, when ML data was available, rated their perceived reliability of the model (see Section 3.6). Hereafter, we refer to these as "red" and "blue", respectively, in the rest of the paper.

A web-based platform was chosen because it enabled remote participation (allowing for broad recruitment across institutions), allowed more precise control over which information was displayed at each tier as well as a more standardized interface that minimized variability in the information presented.

All RCT and ML model data were synthetically generated rather than drawn from real trials or deployed models (see Section 3.3 for details on the data-generating process). This design choice ensured that participants' prior familiarity with specific trials or specific ML systems did not influence their responses. It also allowed fast iteration and conducting of the study given debugging/calibration of real model outputs was not a concern. In real-world settings, clinicians would have access to the identity and reputation of the trial and model, along with additional contextual information such as patient preferences and prior treatment history. We discuss this trade-off between experimental control and ecological validity in Section 5.6. This study was deemed IRB exempt by the Massachusetts Institute of Technology.

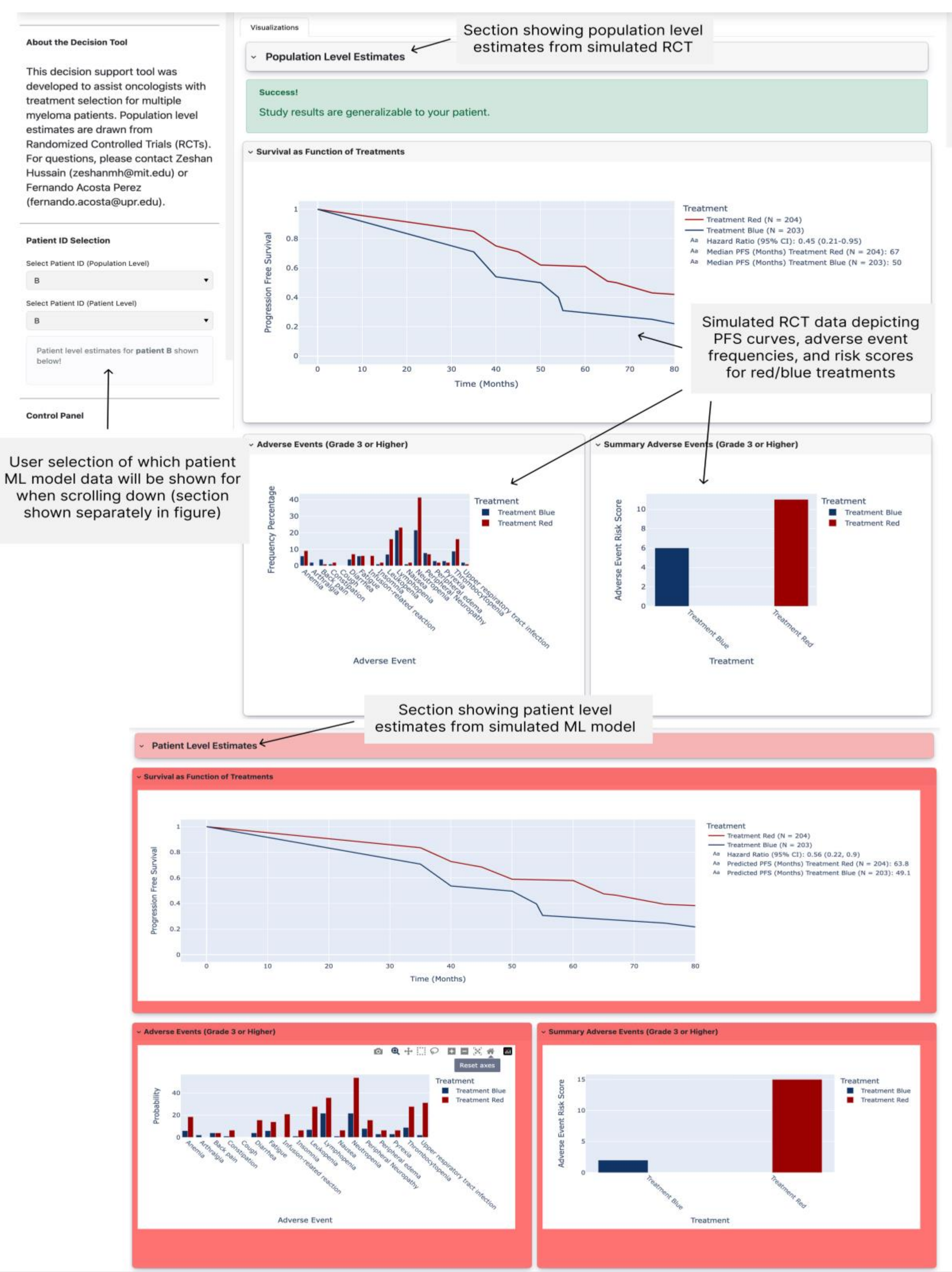

*Figure 1*: Clinical decision support system (CDSS) created for the study, called the “Multiple Myeloma Decision Support Tool”. The left-side panel allowed participants to select one of twelve patient scenarios. The main display area presented two types of data presented on top and bottom: a) population-level data from the simulated RCT, including progression-free survival curves and adverse event frequencies for the overall trial population, and (b) patient-level data from the simulated ML model, including individualized survival predictions and adverse event estimates for the specific patient scenario. Adverse events included symptoms such as cytopenias and infusion-related reactions. The green banner at the top ("Success! Study results are generalizable to your patient") indicated that the patient met the RCT inclusion criteria; in scenarios where the patient did not meet inclusion criteria, this banner was replaced with a warning message (see Appendix File 4). All data were synthetically generated (see Section 3.3), ensuring that participants' prior familiarity with specific trials or models did not influence their responses.

### 3.2 Patient Scenarios and RCT/ML Combinations

Each scenario was based on the same clinical vignette of a 62-year-old man newly diagnosed with MM (ISS Stage II), with four systematically varied clinical variables. These variables were selected in consultation with an MM specialist (co-author AJY) because they are among the most clinically impactful factors that determine both RCT eligibility and treatment outcomes in MM:

*ECOG performance status* (0 vs. 3): Performance status is one of the most common RCT exclusion criteria. ECOG 0 indicates a fully active patient; ECOG 3 indicates a patient capable of only limited self-care.

*Chronic kidney disease* (present vs. absent): CKD affects RCT eligibility (renal impairment is a frequent exclusion criterion) and determines whether the ML model was trained on similar patients.

*Cytogenetic risk profile* (normal vs. high-risk): Cytogenetic abnormalities directly affect prognosis and treatment response, influencing the ML model's survival predictions.

*COPD* (present vs. absent): This comorbidity may affect adverse event risk (e.g. increased risk of pneumonitis in patients with pre-existing airway disease), influencing the ML model's adverse event predictions.

Table 1 presents the full experimental design. The RCT consistently showed a survival benefit with red and comparable adverse events between treatments across all 12 scenarios (since the RCT itself is a population-level result and does not change based on the patient). The ML model intuitively results in differing results based on the patient scenario since it is a function that has a unique output for each unique input.

Although real-world decision-making involves many additional factors (prior treatment history, patient preferences, financial considerations, etc.) beyond these four variables, these were chosen to create a tractable experimental design that captured the most clinically impactful dimensions while keeping the study manageable for participants who evaluated all 12 scenarios.

| **Covariate** | **Values** | **Output** | **Mechanism** |
|---|---|---|---|
| *ECOG* | 0 | Patient meets inclusion criteria of RCT (+RCTi) | ECOG >= 2 is a common RCT exclusion criteria |
| | 3 | Patient does not meet inclusion criteria of RCT (-RCTi) | |
| *History of CKD* | No | ML model was trained on similar patients (+MLtr) | CKD patients under-represented in ML training data |
| | Yes | ML model was not trained on any similar patient (-MLtr) | |
| *Cytogenetic Risk* | Normal Risk | ML model shows benefit with red | High risk => ML predicts no survival benefit with red |
| | High Risk | ML model shows no benefit with red | |
| *History of COPD* | No | ML model shows similar adverse events to RCT | COPD => ML predicts worse AE with red (except J, L) |
| | Yes | ML model shows worse adverse events with red* | |

| | Covariates | | | | Experimental Conditions | | ML Predictions | | |
|---|---|---|---|---|---|---|---|---|---|
| **Scenario** | ECOG | CKD | Cyto Risk | COPD | RCTi | MLtr | **Survival** | **AEs** | **Tier 4** |
| *+RCTi, +MLtr* | | | | | | | | | |
| A | 0 | No | Normal | No | + | + | Benefit with red | Similar to RCT | — |
| B | 0 | No | Normal | Yes | + | + | Benefit with red | Worse with red | — |
| C | 0 | No | High-risk | No | + | + | No benefit with red | Similar to RCT | 4C only |
| D | 0 | No | High-risk | Yes | + | + | No benefit with red | Worse with red | — |
| *-RCTi,+MLtr* | | | | | | | | | |
| E | 3 | No | Normal | No | - | + | Benefit with red | Similar to RCT | 4C + 4NC |
| F | 3 | No | Normal | Yes | - | + | Benefit with red | Worse with red | — |
| G | 3 | No | High-risk | No | - | + | No benefit with red | Similar to RCT | — |
| H | 3 | No | High-risk | Yes | - | + | No benefit with red | Worse with red | — |
| *-RCTi,-MLtr* | | | | | | | | | |
| I | 3 | Yes | Normal | No | - | - | Benefit with red | Similar to RCT | 4C + 4NC |
| J | 3 | Yes | Normal | Yes | - | - | Benefit with red | Worse with blue^ | — |
| K | 3 | Yes | High-risk | No | - | - | No benefit with red | Similar to RCT | — |
| L | 3 | Yes | High-risk | Yes | - | - | No benefit with red | Worse with blue^ | — |

*^ Scenarios J and L belong to the −RCTi, −MLtr condition (CKD=Yes), in which the patient is underrepresented in the ML training data. A model extrapolating outside its training distribution may produce unreliable or inverted adverse event predictions; the worse-AE reversal to blue treatment reflects this possibility. Additionally, if "worse AE" always pointed to red, treatment switching could reflect a label heuristic rather than responsiveness to specific ML predictions; thus, these two scenarios serve as an internal validity check.*
*4C = concordant replication; 4NC = non-concordant replication; — = no Tier 4 data.*
*Table 1*: Combinations of RCT and ML outcomes presented to study participants. Note that RCT consistently showed a survival benefit with red and comparable adverse events between treatments across all 12 scenarios Top: Shows how each covariate maps to its experimental role, Bottom: Shows the resulting 12 scenarios (A--L), grouped by the three RCT-inclusion / ML-representativeness conditions. Excluded scenarios where +RCTi,-MLtr in an attempt to shorten the study and remove less compelling scenarios.

### 3.3 Data Generating Process for Simulated RCT & ML Model

Survival curves and adverse outcomes for the synthetic RCT as well as the inclusion criteria were loosely based on the results from the SWOG S0777 trial [2]. Specifically, we used the PFS curves given in the trial plus added gaussian noise as the survival curves for treatment red and blue in our RCT. We used largely the same adverse event frequencies found in the trial and then consulted with a multiple myeloma specialist to set the frequencies for the remaining adverse events. The trial was effectively "de-identified" with the addition of gaussian noise and synthetic imputation of the adverse event frequencies.

The synthetic ML model outputs were generated using a data-generating process designed to produce clinically plausible, individualized predictions. Survival curves were generated using a Cox proportional hazards framework:

$$PFS_U(t,X) = s_0(t) \cdot \exp\left(\sum_i \beta_i X_i\right) \quad (*)$$

where $U$ denotes the treatment (red or blue), $X$ represents the patient's clinical covariates (Table 1), $t$ denotes a particular time point, and $s_0$ is the baseline survival function, which was set to the progression-free survival curve from the simulated RCT (as above). Intuitively, (*) is shifting the population-level PFS curve from the synthetic RCT upwards or downwards based on the weights, $\beta_i$.

Adverse event frequencies were generated using an analogous function:

$$AE_U(j,X) = AE_j \cdot \exp\left(\sum_i \gamma_i X_i\right) \quad (**)$$

where $j$ indexes a particular adverse event and $AE_j$ is the baseline frequency from the simulated RCT. (**) operates similarly as (*), where the adverse event frequency for a particular patient is adjusted based on a weighted sum of their covariates, $X$. The covariate weights $\beta_i$, $\gamma_i$ were determined in consultation with an MM specialist to produce informative, clinically meaningful individual-level changes in survival and adverse event predictions based on the patient scenarios. Adverse event risk scores for a particular treatment were an unweighted sum of the number of adverse events for which that treatment has a higher probability than the comparator treatment. See Figure 1 for example RCT and ML data. We chose to use simulated rather than real ML model outputs for two reasons. First, it allowed us to precisely control the relationship between RCT and ML predictions across scenarios, enabling systematic comparison of concordant and discordant conditions. Second, it prevented participants from recognizing outputs from specific published models, which could have introduced confounds related to prior model familiarity or reputation.

### 3.4 Tiered Information Approach

Each patient scenario presented information in three to four sequential "tiers", designed to progressively disclose data and assess how additional context changed treatment decisions. In tier 1 (RCT data only), participants viewed population-level survival curves (PFS for red vs. blue) and adverse event bar plots from the simulated RCT in the CDSS. The CDSS also displayed a green "Success" banner when the patient met RCT inclusion criteria (+RCTi), or a warning banner when they did not (-RCTi). After reviewing this information, participants selected a treatment and rated their confidence.

In tier 2 (RCT + ML data), the CDSS added patient-level survival curves and adverse event predictions generated by the simulated ML model. These were displayed below the RCT data, with the RCT outputs labeled as "Population Level" and ML outputs labeled as "Patient Level." Participants again selected a treatment, rated their confidence, and additionally rated their perceived reliability of the ML model.

In tier 3 (RCT + ML data + model quality information), participants received textual information about the ML model's training and validation, presented in the Qualtrics survey. This included 1) whether the patient was well represented in the training data or not, and 2) external validation performance metrics (Appendix Figure 3a). Participants again provided treatment choice, confidence, and reliability ratings.

In tier 4 (replication experiment; scenarios C, E, I only), three scenarios additionally presented the results of a "replication experiment." This procedure proposed to use causal inference methods (target trial emulation [34,35]) to test whether the ML model could replicate the RCT's findings when applied to a matched observational cohort, as an additional means of building trust in the ML model. Participants were told either that the replication was successful (the ML model's predictions matched the RCT results on the observational cohort) or that it failed (Appendix Figure 3b). This tier was designed to examine how additional benchmarking of the ML model influenced physician confidence. Scenarios C, E, and I were selected for Tier 4 because they represented one scenario from each of the three main experimental conditions (+RCTi/+MLtr, -RCTi/+MLtr, -RCTi/-MLtr).

See Appendix File l for the full survey as well as corresponding panels from the CDSS.

### 3.5 Participant Steps

Each study session consisted of the following steps:

A. *Tutorial*: Participants watched a 5-minute instructional video explaining the CDSS interface, the meaning of survival curves and adverse event plots, and the structure of the tiered information approach.
B. *Demographics and background*: Participants completed a brief demographic questionnaire and self-assessed their comfort with managing MM patients, interpreting RCT data, machine learning, and causal inference on a 1-5 Likert scale.
C. *Scenario evaluation*: Participants evaluated all 12 patient scenarios (A-L) in a randomized order. For each scenario, they read the clinical vignette, then progressed through Tiers 1 through 3 (or Tiers 1 through 4 for scenarios C, E, and I), answering treatment choice, confidence, and reliability questions at each tier (see Appendix File 1 for the full questionnaire). Participants accessed the CDSS in a separate browser tab and toggled between the CDSS and the Qualtrics survey.
D. *End-of-study questions*: Participants answered questions about their understanding of the survival curves, the influence of adverse event data on their decisions, and their overall impressions of ML-driven estimates (Appendix File 4).
E. *Exit interview (optional)*: All participants were invited to participate in a 15 minute semi-structured exit interview in which they worked through Scenario K while thinking aloud (see Section 3.6).

The full study session, excluding the optional exit interview, took approximately 30-45 minutes to complete.

### 3.6 Data Collection

At each tier of information, participants were asked to: (1) select a treatment ("Treatment Red" or "Treatment Blue"); (2) rate their confidence in their treatment recommendation on a scale from 1 ("not confident at all") to 10 ("extremely confident"); and (3) when ML data was available (Tiers 2-4), rate the perceived reliability of the ML model predictions on a scale from 1 to 10. We use "confidence" throughout this paper to refer specifically to this self-reported confidence in treatment recommendation.

The treatment labels "Treatment Red" and "Treatment Blue" were chosen as neutral placeholders to avoid using real drug names, which could introduce confounds related to participants' prior experience with specific agents.

Prior to the main study, participants self-assessed their comfort with machine learning and causal inference on a 1-5 Likert scale ("not very comfortable" to "very comfortable"). We note that these single-item self-report measures are inherently coarse and do not distinguish between conceptual familiarity, hands-on experience, and critical evaluation ability. They were used as descriptive characteristics of the sample rather than as predictive variables in our primary analyses. Future studies should consider using validated instruments for measuring ML literacy.

Finally, all participants were invited to participate in a brief semi-structured exit interview lasting 15 minutes. They were asked to verbalize their thought process as they worked through a single scenario (interview protocol in Appendix File 5, COREQ checklist in Appendix File 8). At least one researcher (BDL, ZH) was present during the interview and took field notes.

### 3.7 Recruitment

The study was piloted with two Hematology/Oncology physicians for clarity, length, and clinical relevance (final survey in Appendix File 4). Between January and April 2023, we recruited physicians in Internal Medicine and Hematology/Oncology from two academic institutions via email to participate in our study. At both institutions, trainees in the Hematology/Oncology fellowship program and all attending Hematologists were invited to participate. At one institution, the Internal Medicine residency was also invited to participate. In the U.S. medical training system, residents are physicians completing their initial postgraduate training in a broad specialty (e.g., Internal Medicine), fellows are physicians pursuing subspecialty training (e.g., Hematology-Oncology), and attendings are fully trained physicians who have completed all training requirements. Participants were offered a gift card as an incentive and were permitted to invite others to participate.

### 3.8 Data Analysis

We used descriptive statistics to analyze respondent characteristics. For each of the 12 scenarios, we conducted planned pairwise comparisons using paired *t*-tests: confidence at Tier 1 vs Tier 2, Tier 2 vs Tier 3, and Tier 1 vs Tier 3; perceived reliability at Tier 2 vs Tier 3, and where applicable, confidence and reliability at Tier 3 vs Tier 4. Paired *t*-tests were chosen to assess specific, directional comparisons between adjacent tiers (e.g., "Does adding ML data change confidence?") rather than testing for any overall tier effect. All paired *t*-tests are reported in APA format: *t*(df), *p*-value, with Cohen's *d* (paired) as the effect size measure, where positive/negative *d* indicates higher values at the earlier/later tier.

We also ran McNemar's tests to evaluate changes in treatment selection (red vs. blue) between tiers. Because treatment choice (categorical) and confidence/reliability (continuous) are distinct dependent variables, we applied Bonferroni correction within separate families: 54 paired *t*-tests ($\alpha = 0.05/54 = 0.0009$) and 24 McNemar's tests ($\alpha = 0.05/24 = 0.0021$). In the results section, * denotes significance at the uncorrected $\alpha = .05$ threshold, and *** denotes significance at the Bonferroni-corrected threshold.

Finally, we conducted omnibus one-way repeated-measures ANOVAs for each scenario to test whether confidence (or reliability) differed across tiers. These are reported as $F(df_1, df_2)$, where $df_1$ is the number of tiers minus one and $df_2$ reflects the degrees of freedom for within-subject error. Partial eta-squared ($\eta^2 p$) is reported as an effect size, indicating the proportion of within-subject variability in ratings accounted for by differences across tiers. A power analysis for paired *t*-tests with continuous outcomes indicated a requirement of 34 participants to detect a mean difference of 1.0 with a standard deviation of 2.0 ($\alpha = 0.05$, power = 0.80). Our sample of 32 participants provides 78% power for this effect size.
Analysis was done using the Scipy and Statsmodels Python libraries.

Two researchers (BDL, ZH) used Braun and Clarke's methods [28] to conduct a qualitative thematic analysis of the free text responses and field notes from the exit interviews by first reviewing all responses and independently creating a list of themes to reflect the data, then discussing the themes and iteratively developing a final codebook by consensus. The two researchers used this codebook to independently code the responses, assigning up to two codes per response. The researchers compared their coding assignments and reached consensus for every response. Themes were derived directly from the data and all analysis was conducted using Microsoft Excel.

## 4 RESULTS

A total of 284 physicians were invited to the study and 32 participated, for a response rate of 11.3%. Half were internal medicine residents and half were hematology-oncology fellows and attendings (Table 2).

| Demographic | *N* = 32,(% of *N*) |
|---|---|
| Age | |
| 21-30 years old | 16 (50%) |
| 31-40 years old | 16 (50%) |
| Sex | |
| Male | 23 (72%) |
| Female | 9 (28%) |
| Race | |
| White Caucasian | 22 (69%) |
| Asian | 7 (22%) |
| Hispanic | 2 (6%) |
| Black or African American | 1 (3%) |
| Clinical Role | |
| Internal Medicine Resident | 16 (50%) |
| Hematology Oncology Fellow | 13 (41%) |
| Hematology Oncology Attending | 3 (9%) |
| Clinical Specialty* | |
| General Medicine | 15 (47%) |
| Hematology | 11 (34%) |
| Oncology | 10 (31%) |
| Other | 2 (6%) |
| Background** | |
| Comfort with machine learning | 2.2 +/- 0.8 |
| Comfort with causal inference | 2.1 +/- 1.0 |
| Comfort with RCT survival interpretation | 3.5 +/- 0.9 |
| Comfort with RCT adverse event interpretation | 3.4 +/- 0.8 |

RCT = randomized controlled trial

*Participants could select more than one, thus totals may sum to greater than 100%

**Asked to rate on scale from 1 (“not comfortable”) to 5 (“very comfortable), results recorded as mean +/- standard deviation of rating

***Table 2*: Study cohort**

### 4.1 Quantitative Analysis

Omnibus repeated-measures ANOVAs confirmed a significant main effect of tier on confidence in 10 of the 12 scenarios (Table 1). The two non-significant scenarios were G (F(2,62) = 1.16, *p* = .319, $\eta^2 p$ = .036) and K (F(2,62) = 2.54, *p* = .087, $\eta^2 p$ = .076). In both cases, the ML model disagreed with the RCT on survival but showed similar adverse events; thus, the ML prediction may not have provided a strong enough signal to shift confidence in either direction. All reliability ANOVAs (for scenarios C, E, and I where Tier 4 data were available) were highly significant ($p < .001$), with large effect sizes ($\eta^2 p$ = .356–.500). The planned pairwise comparisons below detail these effects. The full set of experimental results are given in Appendix File 6.

***Patient meets inclusion criteria for RCT and ML model was trained on data that represents the patient well (scenarios A, B, C, D)*** – In scenario A, where ML model results were concordant with RCT results, all 32 participants chose red at Tier 1 & 2, and confidence was the highest observed across all scenarios at Tier 3 (*M = 7.84, SD = 1.19*) and increased from Tier 1 to 3 (*t*(31) = −2.55, *p* = .016, *d* = −0.45*), though individual tier transitions were not significant after correction. In scenario B, where the ML model showed survival benefit with red but worse adverse events, 9 participants (28.1%) switched from their Tier 1 choice (*p* = .046*) but confidence decreased from Tier 1 to 2 (*T1: M* = 7.12, SD = 2.09; *T2: M* = 6.03, *SD* = 1.98; *t*(31) = 3.40, *p* = .002, *d* = 0.60*). It then recovered from Tier 2 to 3 (*t*(31) = −4.76, *p* < .001, *d* = −0.84***). In scenario C, where the ML model showed no benefit with red and similar adverse events to the RCT, 21 participants (66%) switched to blue (*p* < .001***); confidence decreased at Tier 2 (*t*(31) = 3.35, *p* = .002, *d* = 0.59*) and partially recovered at Tier 3 (*t*(31) = −2.70, *p* = .011, *d* = −0.48*). In scenario D, where the ML model showed no benefit with red and worse adverse events, 28 participants (88%) switched to blue (*p* < .001***); confidence decreased at Tier 2 (*t*(31) = 2.24, *p* = .033, *d* = 0.40*) and increased again from Tier 2 to 3 (*t*(31) = −3.21, *p* = .003, *d* = −0.57*). Across scenarios B through D, discordant ML predictions reduced confidence upon disclosure (Tier 2), but model quality information (Tier 3) partially restored it. *Takeaway #1: When both RCT and ML results are statistically applicable (patient meets RCT inclusion criteria [+RCTi], ML model training data represents patient well [+MLtr]), physicians demonstrate high confidence and tend to choose treatments with optimal survival and minimal adverse events, prioritizing survival.*

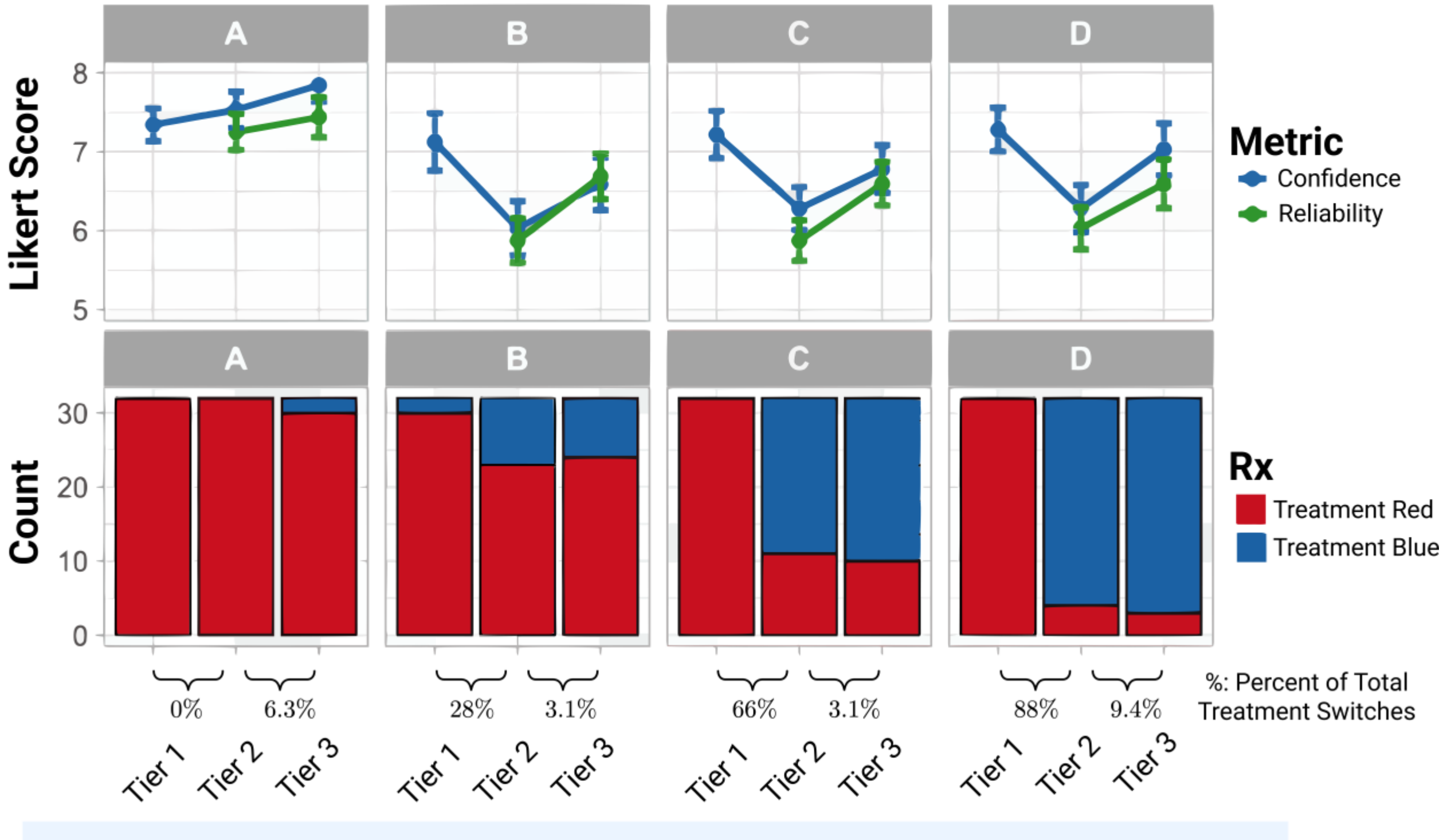

**Scenario A:** ML model shows benefit with red and similar adverse events; +RCTi, +MLtr, +ML/R
**Scenario B:** ML model shows benefit with red but worse adverse events with red; +RCTi, +MLtr, −ML/R
**Scenario C:** ML model shows no benefit with red and similar adverse events; +RCTi, +MLtr, −ML/R
**Scenario D:** ML model shows no benefit with red and worse adverse events with red; +RCTi, +MLtr, −ML/R

*Figure 2*: "Confidence in treatment selection" and "perceived reliability of model predictions" (metrics) as well as treatment counts (Rx) in scenarios A-D as function of tier. In these scenarios, the patient meets inclusion criteria for the RCT (+RCTi) and is well represented in the training data used for the ML model (+MLtr). + or –ML/R refers to the ML model yielding a similar or different result compared to the RCT, respectively. Tier 1 provided RCT results alone, tier 2 added ML model results, and tier 3 added information on the quality of the ML model. Error bars in figure represent standard deviation of metrics across all participants. Note: y-axis ranges are truncated to [5, 8] to emphasize tier-to-tier changes.

***Patient does not meet inclusion criteria for RCT but ML model was trained on data that represents the patient well (scenarios E, F, G, H)*** – In scenario E, where ML model results were concordant with the RCT, confidence increased significantly at each tier transition (T1→T2: $t(31) = -2.79$, $p = .009$, $d = -0.49$*; T2→T3: $t(31) = -4.12$, $p < .001$, $d = -0.73$***), with a large cumulative effect from Tier 1 to 3 ($t(31) = -4.79$, $p < .001$, $d = -0.85$***). In scenario F, where the ML model showed survival benefit with red but worse adverse events, 12 participants (38%) switched to blue ($p = .009$*); confidence decreased at Tier 2 ($d = 0.39$*) and was restored at Tier 3 ($d = -0.69$***). In scenarios G and H, where the ML model showed no benefit with red, a majority of participants switched to blue (G: 56%, $p < .001$***; H: 72%, $p < .001$***). Across these four scenarios, confidence and perceived reliability of the model increased when participants learned that the ML model had been trained on patients like theirs (Figure 3). *Takeaway #2: When there are statistical issues with applying the RCT results [–RCTi] but none with the ML results [+MLtr], there is greater variability in the initial treatment decisions and increased blue selection (compared to scenarios A-D) in subsequent tiers, implying more reliance on the ML results when there was discordance between the RCT and ML data.*

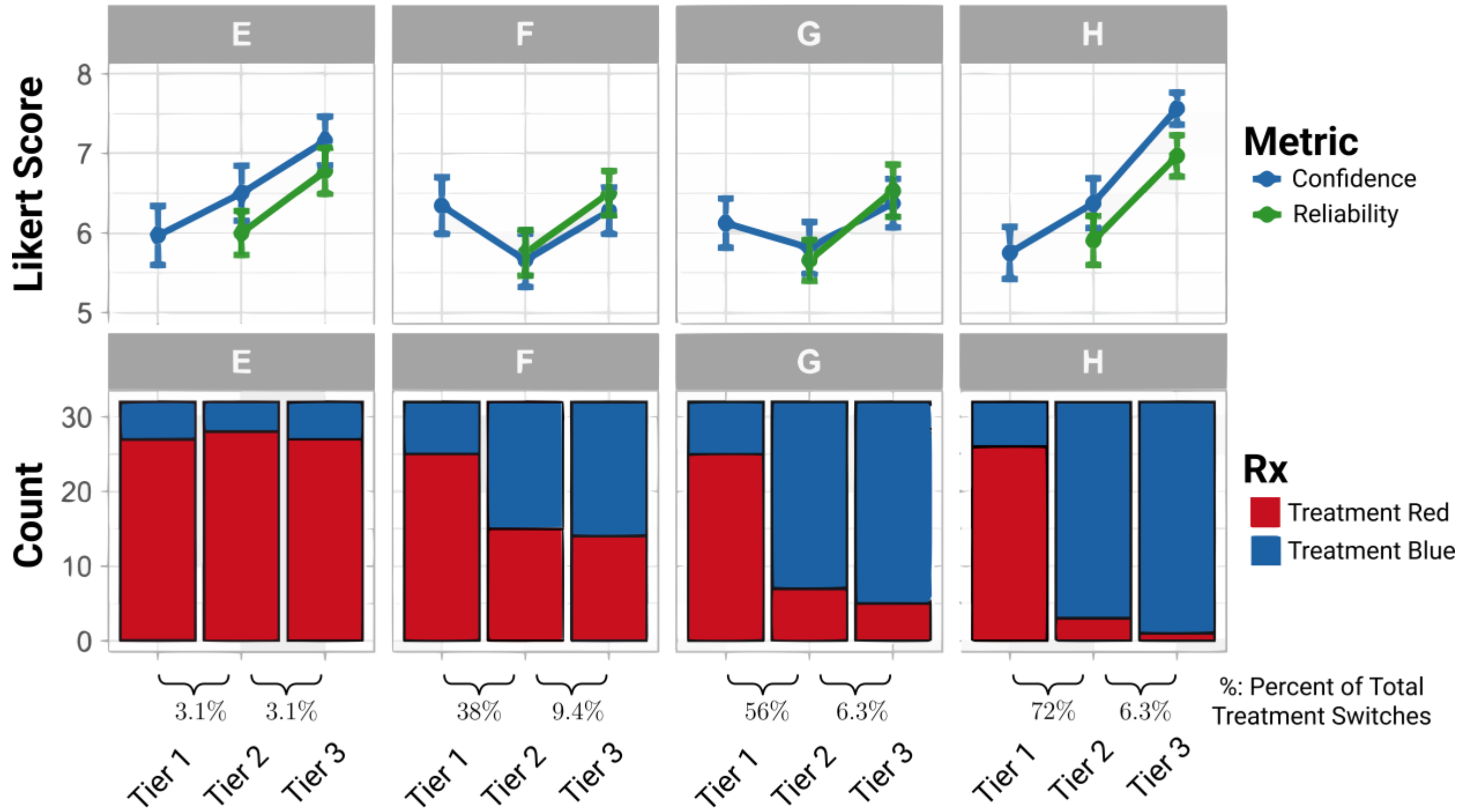

**Scenario E:** ML model shows benefit with red and similar adverse events; −RCTi, +MLtr, +ML/R
**Scenario F:** ML model shows benefit with red but worse adverse events with red; −RCTi, +MLtr, −ML/R
**Scenario G:** ML model shows no benefit with red and similar adverse events; −RCTi, +MLtr, −ML/R
**Scenario H:** ML model shows no benefit with red and worse adverse events with red; −RCTi, +MLtr, −ML/R

*Figure 3*: "Confidence in treatment selection" and "perceived reliability of model predictions" (metrics) as well as treatment counts (Rx) in scenarios E-H as function of tier. In these scenarios, the patient does not meet inclusion criteria for the RCT (–RCTi) but is well represented in the training data used for the ML model (+MLtr). + or –ML/R refers to the ML model yielding a similar or different result compared to the RCT, respectively. Tier 1 provided RCT results alone, tier 2 added ML model results, and tier 3 added information on the quality of the ML model. Error bars in figure represent standard deviation of metrics across all participants. Note: y-axis ranges are truncated to [5, 8] to emphasize tier-to-tier changes.

***Patient does not meet inclusion criteria for RCT and ML model was not trained on data that represents the patient well (scenarios I, J, K, L)*** – In scenario I, where the RCT and ML model results were concordant but neither represented the patient well, >60% chose to treat with red. In scenario J, the ML model showed worse adverse events with blue; 7 participants (22%) switched to red at Tier 2 ($p$ = .023*), and confidence increased substantially ($t(31) = -4.67$, $p < .001$, $d = -0.83$***), but both reversed at Tier 3 (confidence T2→T3: $d = 0.96$***, reliability $d = 0.91$***). In scenario K, the ML model showed no benefit with red and similar adverse events; 23 participants (72%) switched to blue ($p < .001$***), but only 3 (9.4%) switched again at Tier 3 despite learning the model was not trained on similar patients. Confidence dropped from Tier 2 to Tier 3 ($d = 0.62$*) but initial trust in ML predictions persisted in the treatment choice. In scenario L, the ML model showed no benefit with red and worse adverse events with blue; 10 participants (31%) switched at Tier 2 ($p$ = .114) and 9 (28%) at Tier 3 ($p$ = .008*) with half of participants choosing to treat with red (post tier 3). Confidence followed a rise-then-fall pattern (T1→T2: $d = -0.66$***, T2→T3: $d = 0.71$***). Overall, confidence and perceived reliability of the model decreased across all four scenarios when participants learned that the ML model had not been trained on patients like theirs (Figure 5). *Takeaway #3: When neither RCT nor ML model data were statistically applicable [–RCTi, –MLtr], participants showed decreased confidence in the data and less consistency in treatment choices.*

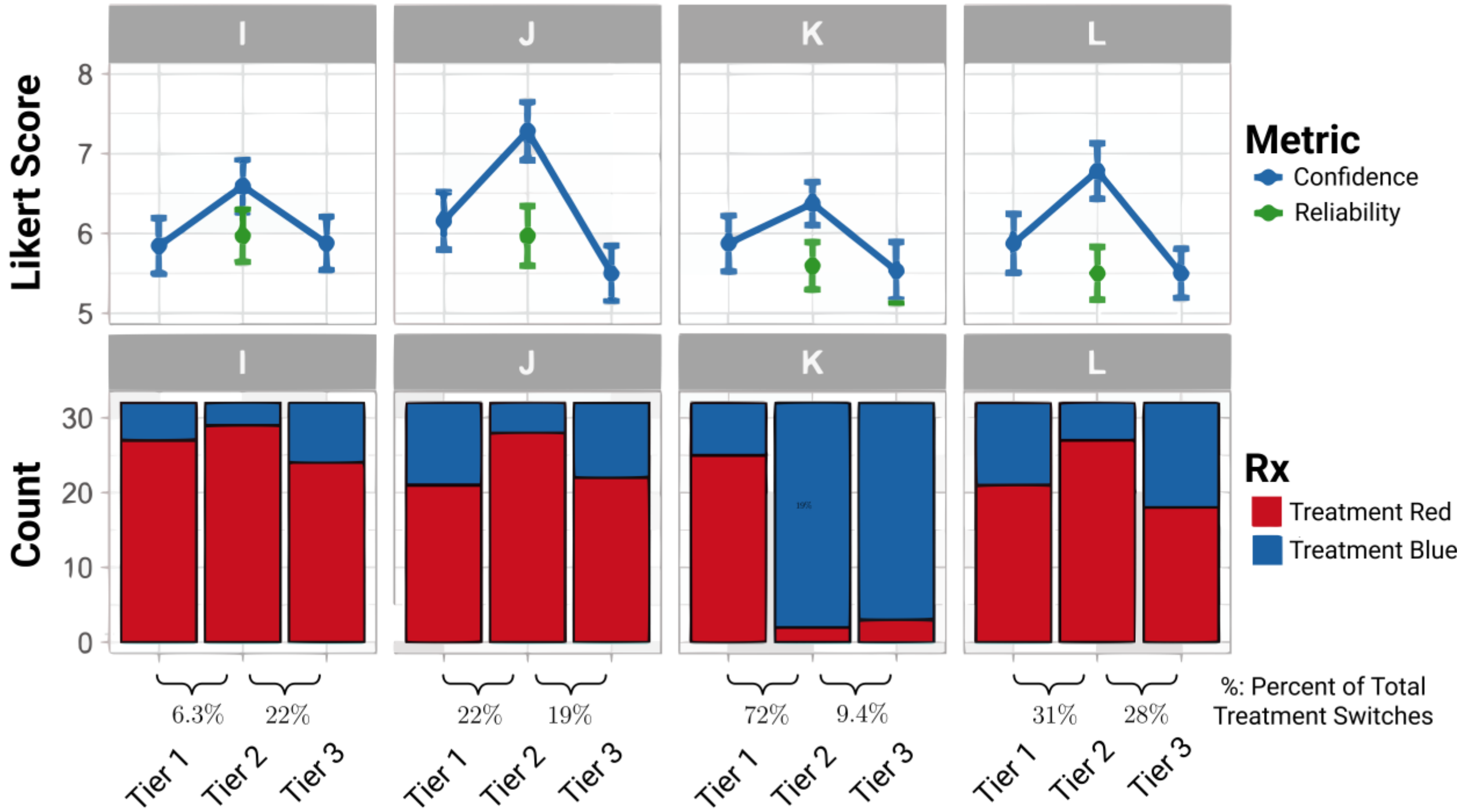

**Scenario I:** ML model shows benefit with red and similar adverse events; −RCTi, −MLtr, +ML/R
**Scenario J:** ML model shows benefit with red but worse adverse events with **blue**; −RCTi, −MLtr, −ML/R
**Scenario K:** ML model shows no benefit with red and similar adverse events; −RCTi, −MLtr, −ML/R
**Scenario L:** ML model shows no benefit with red and worse adverse events with **blue**; −RCTi, −MLtr, −ML/R

*Figure 4*: "Confidence in treatment selection" and "perceived reliability of model predictions" (metrics) as well as treatment counts (Rx) in scenarios I-L as function of tier. In these scenarios, the patient does not meet inclusion criteria for the RCT (–RCTi) and is not well represented in the training data used for the ML model (–MLtr). + or –ML/R refers to the ML model yielding a similar or different result compared to the RCT, respectively. Tier 1 provided RCT results alone, tier 2 added ML model results, and tier 3 added information on the quality of the ML model. Error bars in figure represent standard deviation of metrics across all participants. Note: y-axis ranges are truncated to [5, 8] to emphasize tier-to-tier changes.

### 4.2 Qualitative Analysis of Open-ended Questions

Approximately 82% of the participants responded to the open-ended questions at the end of the study (see E3, E7-E9 in Appendix File 4). Overall, participants prioritized treatments offering better survival outcomes before considering the adverse event data, particularly when the RCT and ML results were applicable to their patient. When asked how they evaluate adverse event data, the majority of participants described looking for specific organ side effects. For example, several participants noted that hematologic toxicities (e.g., neutropenia, thrombocytopenia) were particularly influential, with one participant stating that, "…thrombocytopenia and upper respiratory tract infections were important as … higher risk of patient being admitted." Some discussed evaluating these data in the context of their patient's comorbidities, stating that "risk of neutropenia was pertinent among patients with high-risk features and organ dysfunction, … results may not be generalizable to this population". One participant reported that the total number of adverse events was important in decision-making as well as type of symptom, such as "fatigue, diarrhea, etc.". When asked how they approach treatment selection when their patients do not meet inclusion criteria, the majority reported that adverse events are weighed more heavily. Several described working to balance the benefits of survival outcomes with adverse event data.

When asked how ML results impact clinical decision-making, the majority of participants reported that they find personalized data helpful and that they would incorporate it into their decision-making. “[RCT] results are much harder to generalize for patients who don’t nicely meet the inclusion criteria... with support from ML datasets, we could feel more confident in making more personalized treatment plans and decisions with our patients,” one participant wrote. Several reported that concordant results between the RCT and ML model increased confidence. For example, as one participant noted, “When they were in agreement, it was a nice confirmation. When there was discrepancy, [ML data] was easy to throw out.” A few reported that ML model estimates are not helpful, with one participant stating, “They may bias towards increasing confidence but [they] rarely changed my mind.”

### 4.3 Replication Experiment: Quantitative & Qualitative Analysis

Participants were presented with replication experiment results (tier 4 data) in scenarios C, E, I. In all three scenarios, when the replication experiment was successful (ML model results matched RCT results on a matched observational cohort), confidence in treatment and perceived reliability of the model increased. In all three scenarios, when the replication experiment failed, confidence in treatment and perceived reliability of the model decreased. These changes were statistically significant in scenario E, where the patient was not RCT-eligible but was well represented in the model training data (Appendix File 6).

Participants were also asked to describe the replication procedure (tier 4) in their own words. Of the thirty-two participants, twenty were unsure or provided an incorrect definition. However, the remaining participants were able to do so, stating, “Confirming the ML model was reliable” or “Verifying reproducibility.” Several (n = 7) also correctly described the procedure as “simulating” an RCT using observational data.

### 4.4 Exit Interview; Qualitative Analysis of Scenario K

Six of the 32 participants (18.8%) self-elected to participate in a semi-structured exit interview focused on Scenario K (see Appendix File 5 for the interview protocol), the purpose of which was to capture deeper qualitative insight into clinicians’ reasoning processes that could not be captured through the structured survey alone. Three were Internal Medicine residents, two were Hematology and Oncology fellows, and one was an attending, a sample representative of the larger participant group. The following findings should be interpreted as exploratory given the small sample size.

**Clinicians vary in how they use RCT data to make treatment decisions.** Participants described evaluating performance status and its reversibility, comorbidities and their effects on drug metabolism, as well as adverse events and their perceived impact on quality of life. Some focused on performance status — "I worry about performance status" (P1) — while others emphasized drug-comorbidity interactions: "Looking at this drug, will there be an issue when the patient has CKD?" (P3). One attending took a more aggressive stance: "ECOG 3 is likely related to their disease, so I want to be aggressive with treatment" (P6). Approaches to treating patients who did not meet RCT inclusion criteria also varied, from cautious (“It has to be a risk benefit discussion [with your patient]," P2; "You have to think, what are the patient preferences?” P3) to pragmatic (“We routinely use treatments for patients who do not meet inclusion criteria, so I don't worry about that so much,” P6).

**Clinicians trust ML predictions even before receiving information about model quality.** All six interviewees opted for the treatment that the ML model recommended and demonstrated a willingness to incorporate ML predictions into their decision-making before receiving information about model quality. Statements ranged from measured (“I would have some trust in the data,” P1; “I'm assuming the model is taking something specific about the patient into account,” P5) to trusting (“If I take it at face value, assume it's a perfect model, my confidence is high,” P6).

**Clinicians vary in how they interpret information about ML model quality.** When shown the finding in scenario K that most of the 32 participants had chosen treatment blue despite learning that the ML model had not been trained on patients like theirs, none of the participants expressed surprise. Several shared that they also felt conflicted in their treatment choice: “This knocked down my confidence... There's no good, right answer” (P2). One participant noted that “ML models in general do not seem to do as well extrapolating to things outside their training sets” (P3). In contrast, another participant appreciated being told the lack of support in the ML training data: “I thought of [information that the ML model was not trained on similar patients] as a disclaimer and not a big deal since it had good performance” (P4); “Even though the training dataset wasn't representative, it still performed well” (P5). Another appreciated the transparency: “It gives me more confidence because it's being honest about it:” (P4).

**Multiple variables affect real-life decision-making, and this study is a valuable exercise.** Several interviewees noted that different levels of clinical training may affect how providers integrate ML data: “People pay attention to, hey, this model performed really well... people don't really think about ML or the specific issues with ML. People in medicine are not trained to do that” (P3). Others pointed to discipline-specific risk tolerance: “I wonder if a Hematology Oncology fellow would choose differently; people tend to trust their own expertise” (P5); “There's a difference in risk tolerance. I bet the Internal Medicine doctors in your study saw neutropenia and were worried, but the hematologists said, 'no problem'” (P6). All participants reported that the study was a valuable thought exercise.

## 5 DISCUSSION

In our study evaluating how clinicians combine RCT data with ML data to make treatment decisions, four key themes emerged. First, self-reported confidence in treatment recommendation was highest when RCT and ML findings were concordant. Second, clinicians favored survival outcomes regardless of evidence source. Third, clinicians were likely to follow ML estimates even before information about how the model was trained or validated was given. Fourth, most participants could not articulate the model quality procedures they had just used to adjust their decisions.

When the RCT and ML model both represented the patient well and their outcomes agreed, clinicians demonstrated the highest confidence and treatment consistency. When findings were discordant, a substantial proportion of participants switched to the treatment supported by the ML model. In scenario C (RCT shows red benefit; ML shows no red benefit – adverse events were similar), 66% of participants switched to blue ($p < .001$). In scenario D (RCT shows red benefit; ML shows no benefit and worse adverse events w/ red), 88% switched ($p < .001$). This result indicates clinicians may prioritize survival outcomes over other types of information, consistent with prior reports in the cancer literature [29]. In our study, this preference was observed whether the benefit was estimated by the RCT or the ML model, and before participants received information on how the model was trained and validated. In scenarios where the patient did not meet RCT inclusion criteria, participants were more likely to adhere to ML predictions. The majority continued to do so even when neither the RCT nor the ML training data represented the patient well. Interestingly, in exit interviews focused on this scenario, participants tended to favor the RCT-supported treatment when told the ML model was not trained on patients like theirs, suggesting observer effects may have influenced behavior.

That clinicians followed ML predictions before reviewing model quality information raises the question of whether these responses reflect appropriate reliance. As mentioned, in discordant scenarios, most participants shifted toward the ML-supported treatment at Tier 2, before any model quality information was disclosed. This pattern is consistent with automation bias, which persists even among expert users [47] and represents a failure of appropriate reliance [50]. At the same time, our qualitative data reveal a countervailing pattern. One participant noted that when RCT and ML data disagreed, “[ML data] was easy to throw out”, which is consistent with algorithm aversion [48], and suggests that some clinicians default to the familiar evidence hierarchy in which RCTs occupy the highest tier. Quantitative evidence of over-reliance co-existing with qualitative signals of algorithm aversion across participants in our study mirrors the broader literature [25, 26], but the presence of a competing evidence source (RCT data) may amplify algorithm aversion by giving clinicians a concrete alternative to fall back on. This is a dynamic not captured in single-source studies.

Against this backdrop as well as recent explainability research revealing increasing reliance regardless of correctness [52], our Tier 3 results add more nuance. They show that confidence increased when participants learned the model was trained on similar patients (+MLtr) and decreased when it was not (−MLtr), and in scenarios where participants were presented with a causal inference approach to validating the model (Tier 4), confidence dropped significantly when the replication procedure failed [34, 35]. This suggests that specific, substantive disclosures about training data representativeness can promote calibrated trust in ways that generic explanations do not. CDSS designers might consider cognitive forcing functions [42] that require review of model quality/transparency before predictions are displayed. Our tiered design models this pattern.

The risk of both over-reliance and aversion is compounded by a gap between behavioral response and comprehension. Twenty of 32 participants (62.5%) could not correctly describe the replication procedure (Section 4.3), despite having just used its results to adjust their confidence and reliability ratings, indicating a gap between appropriate behavioral response and true understanding. This challenge is compounded by heavy workloads and time pressure in clinical settings.

Some have called for using RCTs to evaluate models before deployment [30], but this approach is limited by cost and time [31, 32], and the development of ML models will likely outpace our ability to conduct robust trials. Rather than expecting individual clinicians to appraise each model, a more sustainable approach would combine regulatory pre-screening, interface designs that surface model quality in actionable ways (e.g., out-of-distribution warnings, conveying predictive uncertainty directly [36, 43]), and integration of AI literacy into medical education [33]. Case studies of challenging scenarios such as those presented in this study may help clinicians develop a mental model for how they would manage AI in clinical practice.

How ML model results are displayed may also influence users. In our study, the labeling of RCT data as "Population Level" and ML predictions as "Patient Level" possibly primed participants to view ML outputs as more personally relevant. Additionally, the sequential tier structure may have anchored initial decisions on RCT data. These concerns are consistent with research showing that presentation format significantly affects clinical interpretation of AI outputs [41], and motivated our choice to present ML predictions as outcome estimates rather than treatment recommendations [27]. Future work should systematically vary interface design elements to isolate their effects on clinician behavior.

### 5.1 Limitations

We had a response rate of 11.3% (32 of 284 invited), though this is not dissimilar to other web-based study response rates among clinicians [39]. All participants were younger than 40 years old, and 91% were in training (residents or fellows). Younger clinicians may view ML more favorably [40, 44], and our findings may not generalize to older, more experienced practitioners or to non-academic settings.

In addition, our experimental setting prioritized internal validity and experimental control to establish baseline findings at the expense of ecological validity. We used simulated rather than real RCT and ML data, which prevented participants from drawing on prior knowledge of specific trials or models but ensured that treatment decisions were driven by the experimental manipulations rather than pre-existing familiarity or opinions. The forced binary treatment choice (red treatment vs. blue treatment) removed the option to defer or seek additional information, which may not be realistic in practice. However, this decision did allow us to cleanly observe which direction clinicians leaned as new information was disclosed, and participants' confidence ratings provided a continuous proxy for deferral intent. We presented a single simulated RCT and four clinical variables to isolate the key experimental conditions (±RCTi, ±MLtr, survival concordance, adverse event concordance). Statements about RCT applicability and ML training data representativeness were pre-provided to standardize the information each participant received. In practice, these are non-trivial assessments that clinicians must make independently, and future work should examine how this additional judgment layer affects decision-making. Our study isolates the cognitive dimensions of clinician–AI interaction, though deployment studies that incorporate real patient data, multidisciplinary discussion, time pressure, and other contextual factors [51] are needed to understand how these dynamics manifest in practice.

In our exit interviews, only 6 of 32 participants (18.8%) elected to participate, and data saturation was not reached; findings from the interviews should be considered exploratory. We took field notes rather than audio recordings due to the brief nature of the interviews, which may have introduced bias in capturing participants' statements. Further qualitative studies that include a more heterogeneous group of clinicians can help us better understand how RCT and ML data are used in clinical decision-making.

## ACKNOWLEDGMENTS

The authors have no conflicts of interest to disclose. This research was generously supported by an ASPIRE award from The Mark Foundation for Cancer Research. ZH was additionally supported by the National Institutes of Health under Award Number F30CA268631. ZH and BDL had full access to all the data in the study and take responsibility for the integrity of the data and the accuracy of the data analysis. Data is available upon request. Manuscript editing and statistical analysis code were assisted by Claude Code (Anthropic).

## A. APPENDICES

### Appendix File 1: Questionnaire to track treatment recommendations and perceived confidence and reliability

**Patient scenario X**

A 62-year-old man with a history of… (Appendix File 2)

**Tier 1: RCT Data (Population Level)**

After reading the scenario above, go to the web tool and select Patient X in the "Select Patient ID (Population Level)" box. Look only at the population level estimates (note: do not look at the patient level estimates!).

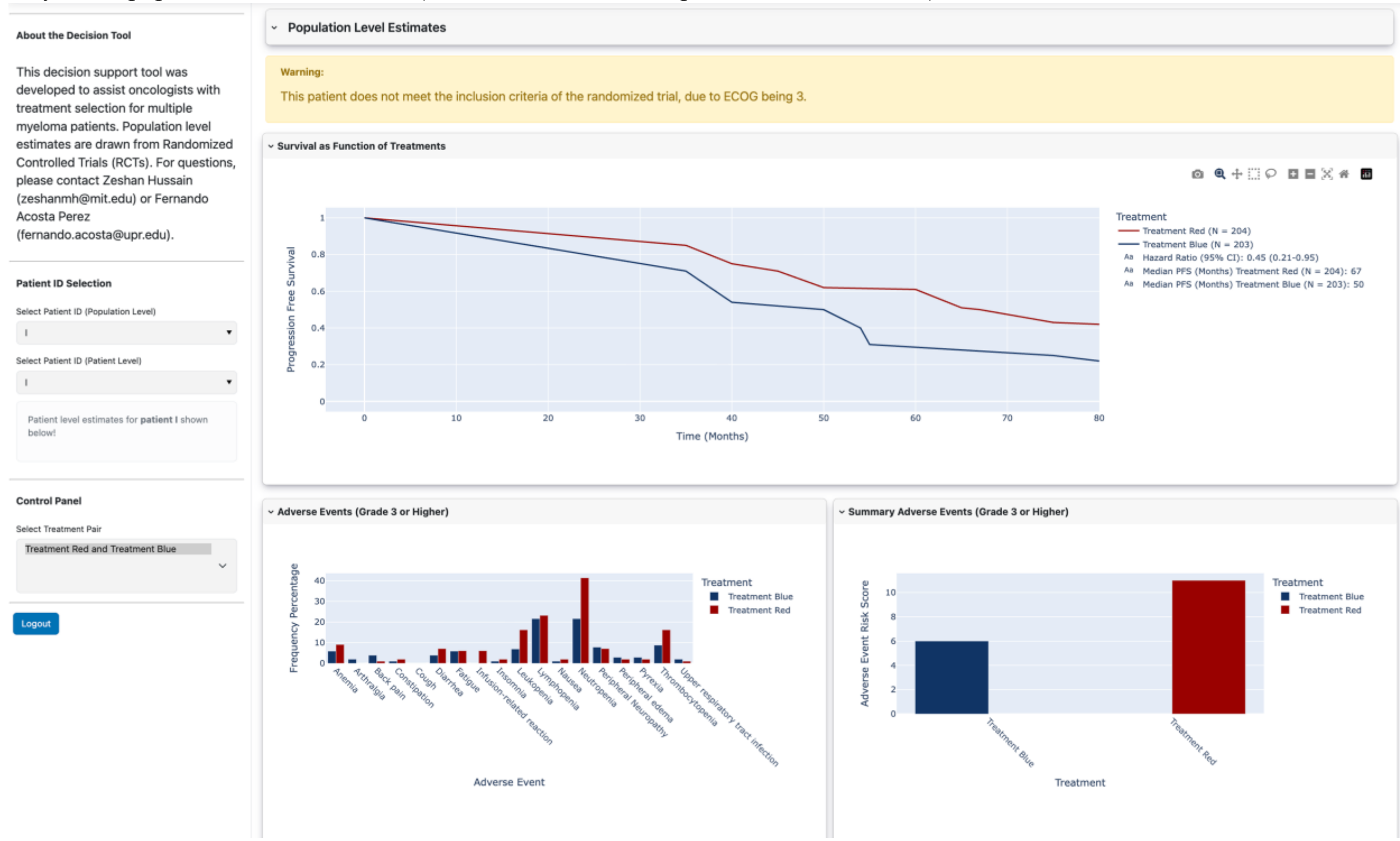

After analyzing the data in the web tool, answer the Patient X – Tier 1 questions below.

What is your treatment recommendation?

- ⇒ Treatment Red
- ⇒ Treatment Blue

How confident are you in your treatment recommendation after seeing Tier 1 data (i.e. population level data only)?

1 2 3 4 5 6 7 8 9 10

1 is "not confident at all" and 10 is "extremely confident"

**Tier 2: RCT Data (Population Level) + ML Data (Patient Level)**

Now, go back to the web tool and look at the patient level estimates for Patient X by selecting "X" in the "Select Patient ID (Patient Level)" box. In addition, you may still look at the population level estimates if you wish; the population level estimates will not change from patient to patient.

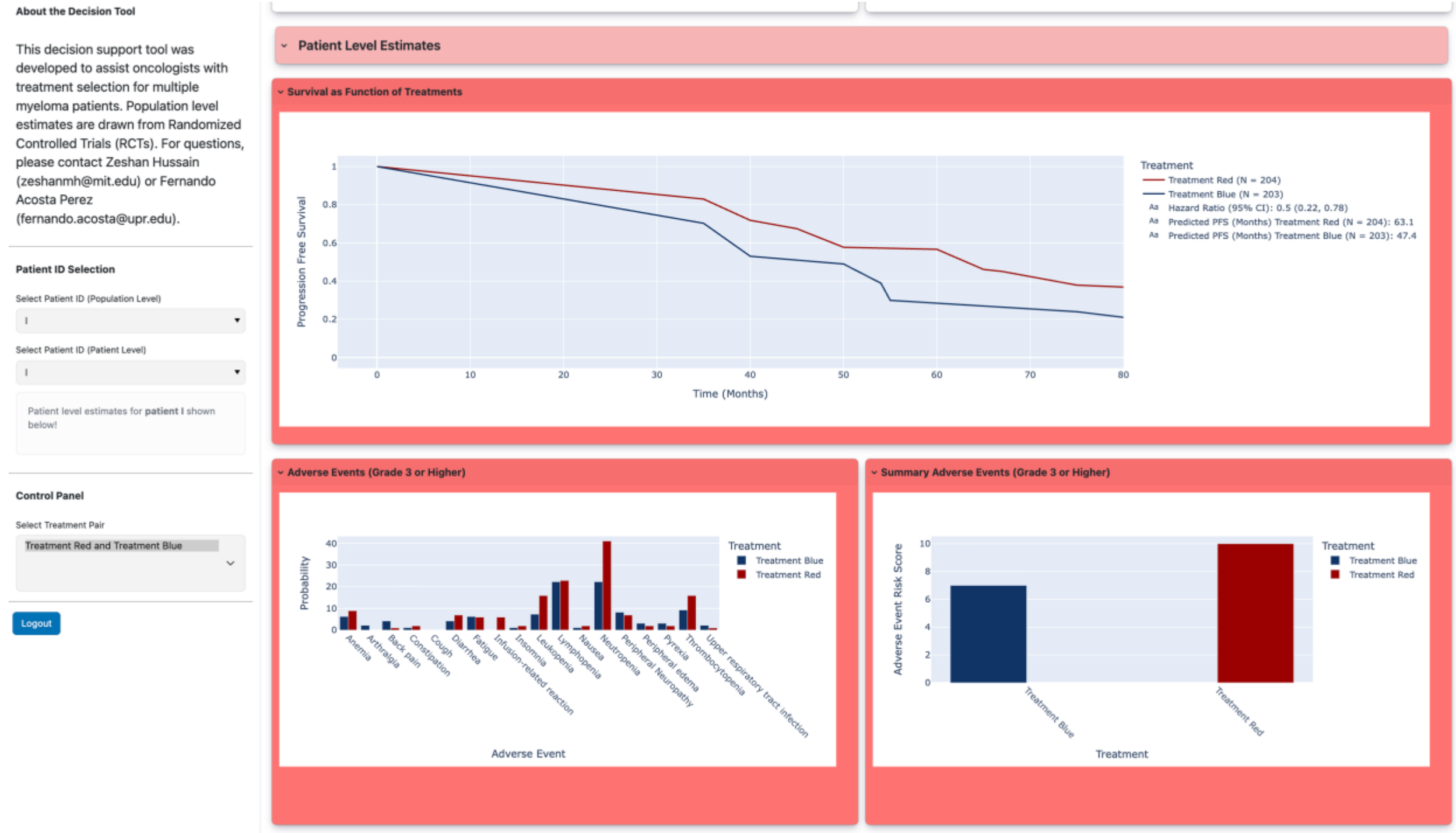

After analyzing the data in the web tool, answer the Patient X – Tier 2 questions below.

What is your treatment recommendation?

⇒ Treatment Red

⇒ Treatment Blue

How confident are you in your treatment recommendation after seeing Tier 2 data (i.e. population level data + patient level data)?

1 2 3 4 5 6 7 8 9 10

1 is "not confident at all" and 10 is "extremely confident"

How reliable do you think the patient level estimates of survival curves and adverse events are, predicted from a machine learning model, after seeing Tier II data?

1 2 3 4 5 6 7 8 9 10

1 is "not confident at all" and 10 is "extremely confident"

Now suppose you are given additional "contextual" information about the machine learning models used (shown below).

**Tier 3: Details about machine learning models**

(Appendix Figure 3a)

A

## Details about the machine learning models

**Data**

**Patient B** is well represented in the model training data. Patients similar to **Patient B** in the training data were equally likely to receive treatment red as they were to receive treatment blue.

**Modeling and Validation**

We trained survival models and adverse event models on the treatment red and blue groups. On an external dataset of multiple myeloma patients from several institutions, the concordance index of both survival models was 0.91 +/- 0.01 and the average ROC-AUC of both adverse event models was 0.92 +/- 0.02.

*(Note: The maximum value of the concordance index and ROC-AUC is 1.0, indicating a very predictive model. Random predictions would yield a ROC-AUC and concordance index of 0.5)*

Using the information above as well as the data given to you so far in the web tool, answer the Patient X – Tier 3 questions below.

What is your treatment recommendation?

⇒ Treatment Red
⇒ Treatment Blue

How confident are you in your treatment recommendation after seeing Tier 2 data (i.e. population level data + patient level data) and Tier 3 data (i.e. population level data + patient level data + validation of ML model used to derive patient level data)?

1 2 3 4 5 6 7 8 9 10
1 is "not confident at all" and 10 is "extremely confident"

How reliable do you think the patient level estimates of survival curves and adverse events are, predicted from a machine learning model, after seeing Tier 2 and Tier 3 data?

1 2 3 4 5 6 7 8 9 10
1 is "not confident at all" and 10 is "extremely confident"

**Tier 4: RCT Data (Population Level) + ML Data (Patient Level) + Extended Context**
Now, you are given further details about the machine learning models below.

(Appendix Figure 3b)

B

## Additional details about the machine learning models

**Replication of RCT**

For additional validation, we ran the following procedure: using our machine learning models, we tried to replicate the results of the RCT by running a "pseudo-experiment" on an observational cohort that matches the RCT cohort.

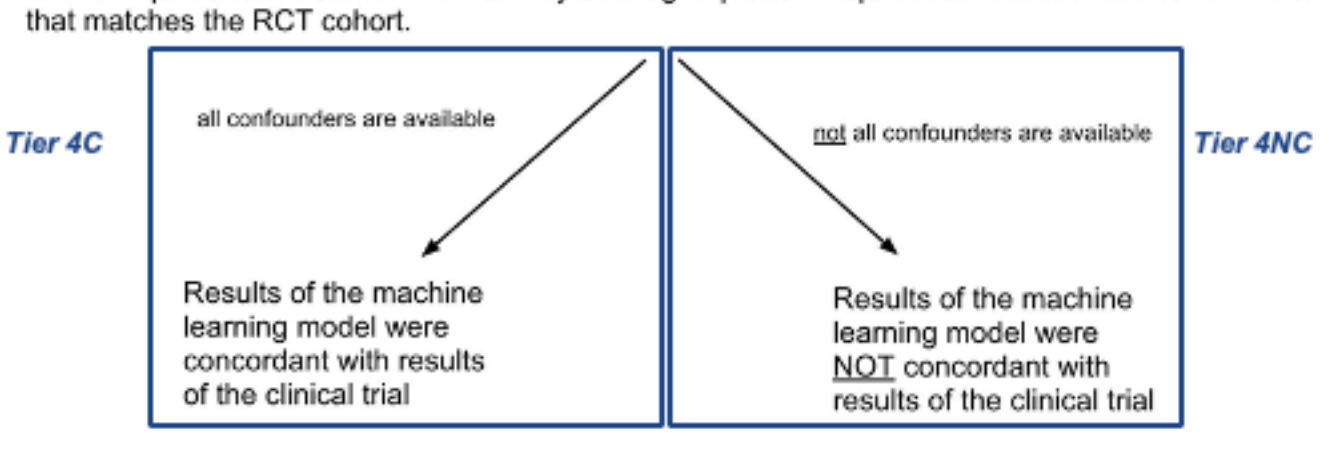

Given all the data so far and the additional information above, answer the Patient X – Tier 4 questions below.

What is your treatment recommendation?

⇒ Treatment Red
⇒ Treatment Blue

How confident are you in your treatment recommendation after seeing Tier 2 data (i.e. population level data + patient level data), Tier 3 data (i.e. population level data + patient level data + validation of ML model used to derive patient level data) and Tier 4 data (i.e. replication)?

1 2 3 4 5 6 7 8 9 10
1 is “not confident at all” and 10 is “extremely confident”

How reliable do you think the patient level estimates of survival curves and adverse events are, predicted from a machine learning model, after seeing Tier 2, Tier 3, and Tier 4 data?

1 2 3 4 5 6 7 8 9 10
1 is “not confident at all” and 10 is “extremely confident”

**Appendix File 2: Changing clinical variables in the patient scenarios**

| **Variable** | **Value** | **Output** |
|---|---|---|
| *ECOG* | 0 | Patient meets inclusion criteria of RCT |
| | 3 | Patient does not meet inclusion criteria of RCT |
| *History of CKD* | No | ML model was trained on patients similar to current patient |
| | Yes | ML model was not trained on any patient similar to current patient |
| *Cytogenetic Risk* | Normal Risk | ML model shows benefit with red |
| | High Risk | ML model shows no benefit with red |
| *History of COPD* | No | ML model shows similar adverse events to RCT |
| | Yes | ML model shows worse adverse events with red* |

A 62-year-old man with a history of [COPD and/or CKD] and nonalcoholic fatty liver disease (NAFLD) presents to doctor for routine check up and is found to have mildly elevated liver function tests (LFTs). As part of his workup, qua immunoglobulins are sent and are found to be low. Next week, his LFTs have normalized and further labs show:

WBC: 4.6 K/uL (within normal range)
Hemoglobin: 10.1 g/dL (abnormal)
Platelet count: 203 K/uL (within normal range)
Creatinine: [1.0 mg/dL (within normal range) / 1.5 mg/dL (abnormal but at his baseline)]
Albumin: 4.4 g/dL (within normal range)
Calcium: 9.5 mg/dL (within normal range)

Serum protein electrophoresis: A monoclonal IgG Kappa is present based on immunofixation. An abnormal band is prese gamma region, roughly 3,574 mg/dL (3.5 g/dL) of total protein (abnormal).

Free Kappa: 60.2 mg/L (abnormal)
Free Lambda: 25.8 mg/L (within normal range)
Free Kappa/Lambda ratio: 2.33 (abnormal)

A PET scan shows a lytic lesion in the right tibia measuring 2.0cm that is FDG avid, concerning for multiple myeloma.

A bone marrow biopsy shows plasma cells occupying 70% of the bone marrow core. Further characterization with cyto and FISH show a [normal risk / high risk] profile. He is a Stage II based on the International Staging System (ISS).

The patient is referred to your hematology clinic. He reports [no symptoms and is completely independent at home (EC
leg pain and needs significant help at home (ECOG 3)]. He is eager to get started on treatment. What medication do yo
start for treatment?

*Note that scenarios J & L show worse adverse events with the blue treatment

**Appendix Figure 3a: Tier 3 information on the machine learning data**

A

Details about the machine learning models

**Data**

**Patient B** is well represented in the model training data. Patients similar to **Patient B** in the training data were equally likely to receive treatment red as they were to receive treatment blue.

**Modeling and Validation**

We trained survival models and adverse event models on the treatment red and blue groups. On an external dataset of multiple myeloma patients from several institutions, the concordance index of both survival models was 0.91 +/- 0.01 and the average ROC-AUC of both adverse event models was 0.92 +/- 0.02.

*(Note: The maximum value of the concordance index and ROC-AUC is 1.0, indicating a very predictive model. Random predictions would yield a ROC-AUC and concordance index of 0.5)*

Additional information about the ML models was displayed in the Qualtrics survey.

*Appendix Figure 3a*: All patient scenarios included tier 3 data, which described how the ML model was trained and validated, and specifically whether the patient was well represented in the training data.

**Appendix Figure 3b: Tier 4 information on the machine learning data**

B

Additional details about the machine learning models

**Replication of RCT**

For additional validation, we ran the following procedure: using our machine learning models, we tried to replicate the results of the RCT by running a "pseudo-experiment" on an observational cohort that matches the RCT cohort.

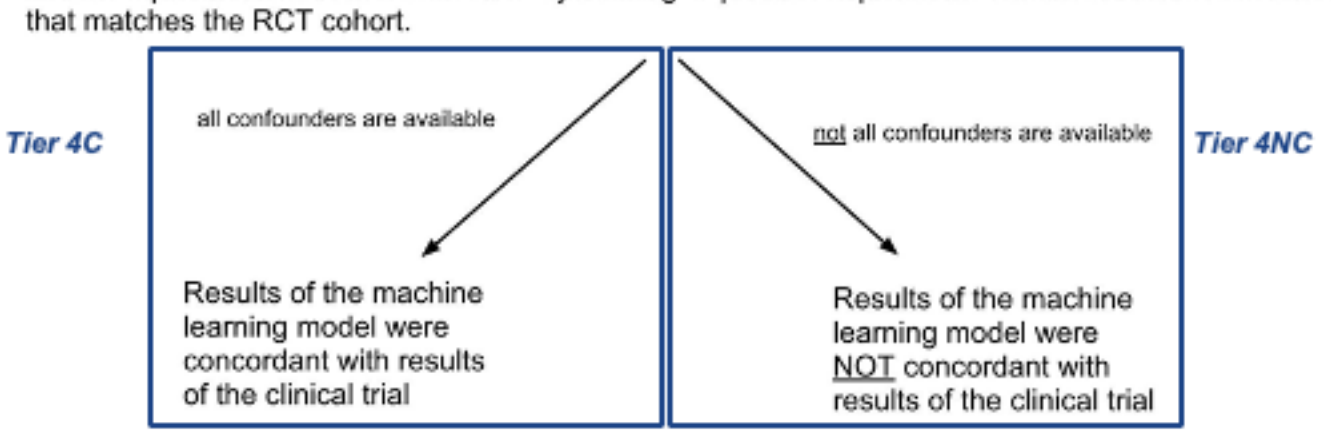

*Appendix Figure 3b*: Three scenarios also included tier 4 data, which described a replication experiment in which the ML model was tested on an observational cohort that matched the RCT cohort.

**Appendix File 4: Study Questions**

Welcome to the Multiple Myeloma Decision Support Tool (MM-DST) user study! This form allows you to enter your medication recommendations for each patient. For each scenario, you will enter your choice of medication as well as your confidence after each "tier" of information is made available to you.

Before beginning, please watch the following five minute tutorial video: https://youtu.be/xguWpxUQmis

Your participation is voluntary and your answers are anonymous. This study meets the criteria for IRB exemption as determined by MIT (E-4559, Decision Support Tool for Treatment Selection in Multiple Myeloma). If you have questions about your rights participating in research or would like to speak with someone independent from the research team, please contact MIT COUHES at couhes@mit.edu. For further questions about the study, please contact the PI at dsontag@csail.mit.edu.

**To begin, we will ask a few demographic questions for post-study analysis. Your answers will be anonymous.**

What is your age range?

- 21-30 years old
- 31-40 years old
- 41-50 years old
- 51-60 years old
- 61 years or older

What gender do you identify with?

- Male
- Female
- Other

What is your race/ethnicity? (Check all that apply)

- American Indian or Alaska Native
- Asian
- Black or African American
- Native Hawaiian or Pacific Islander
- White
- Hispanic
- Other

What is your level of training?

- Resident
- Fellow
- Attending (out of training <10 years)
- Attending (out of training ≥10 years)

What is your specialty? (Select all that apply)

- General Medicine (Hospitalist or PCP)
- Hematology
- Oncology
- Other

**For the following questions, 1 is "not very comfortable" and 5 is "very comfortable"**

How comfortable are you with managing the care of a multiple myeloma patient?

1 2 3 4 5

How comfortable are you interpreting adverse event information found in Randomized Controlled Trials (RCTs)?

1 2 3 4 5

How comfortable are you with machine learning?

1 2 3 4 5

How comfortable are you with causal inference?

1 2 3 4 5

**End of study questions**

(E1) Did you understand what the survival curves show? (1 is "no, not really" and 10 is "yes, completely")

1 2 3 4 5 6 7 8 9 10

(E2) To what extent did the adverse event bar plots inform your treatment selection? (1 is "not at all" and 10 is "very much so")

1 2 3 4 5 6 7 8 9 10

(E3) What adverse events, if any, were important for making a treatment decision?

Rate the following statements from 1 (*no effect on decision-making*) to 10 (*high effect on decision-making*). As a reminder of each of these components, a screenshot is included below.

(E4) To what extent did the "Data" portion of the machine learning model details affect your treatment decision-making and/or your confidences?

1 2 3 4 5 6 7 8 9 10

(E5) To what extent did the "Modeling and Validation" portion of the machine learning model details affect your treatment decision-making and/or your confidences?

1 2 3 4 5 6 7 8 9 10

(E6) To what extent did the "Replication of RCT" portion of the machine learning model details affect your treatment decision-making and/or your confidences?

1 2 3 4 5 6 7 8 9 10

(E7) In your own words, what was the replication procedure doing, at a high level?

Warning:

This patient does not meet the inclusion criteria of the randomized trial, due to ECOG being 3.

(E8) What was your approach for treatment selection when the study results were not applicable to your patient (e.g. you got a statistical warning akin to the screenshot below)?

(E9) Overall, how do you think machine learning-driven, personalized-level estimates of your patient's outcomes impact clinical decision making?

| Question | *Mean +/- Standard Dev. (N=32)* |
|---|---|
| E1 | 8.5 +/- 1.6 |
| E2 | 7.0 +/- 2.0 |
| E4 | 6.6 +/- 1.6 |
| E5 | 5.7 +/- 2.0 |
| E6 | 5.7 +/- 2.5 |

*see above for scales

*Appendix Table 2*: End of survey question results

**Appendix File 5: Interview protocol for exit interview and single scenario analysis**

*Introduce yourself, state your affiliation and role on the research team*.

Thank you for agreeing to take part in this interview. Your participation is voluntary and you may stop the interview at any time. I expect this discussion to last about 15 minutes. We are using this time to get more information about a specific scenario in the study you completed. There are no right or wrong answers. We want to understand how participants are thinking through this scenario. As you work through the scenario, please "think aloud" so that I can understand your thought process. We'll use the term "RCT" to refer to randomized controlled trials and the term "ML" to refer to machine learning.

Question prompts:

**For Tier 1 (RCT data)**

1. How do you interpret the data from this RCT?
   *If participant mentions that their patient does not meet inclusion criteria*,

   1a. How do you think about data from RCTs when your patient does not meet inclusion criteria?

2. What factors are you weighing when choosing a treatment option?
3. How are you weighing the side effects?
4. Why are you choosing that confidence level?

**For Tiers 2 and 3 (ML data)**

1. How do you interpret the data from this ML model?

   *If participant mentions that their patient does not meet inclusion criteria*,
   1a. How do you think about data from ML models when your patient does not meet inclusion criteria?

2. What factors are you weighing when choosing a treatment option?
3. How are you weighing the side effects?
4. Why are you choosing that confidence level?
5. Why are you choosing that level of perceived reliability?

**Finally:**

1. How do you compare the RCT results to the ML results?
2. We found that the majority of participants choose to switch to the blue treatment after seeing the ML data and context (show them ***Scenario K results*** document). Why do you think that is?

Helpful probes:

- Can you talk more about that?
- Help me understand what you mean.
- Can you give an example?

**Appendix File 6: Experimental results for all scenarios**

Full results for all scenarios are shown below. The first table shows the repeated-measures ANOVA to test the omnibus null hypothesis within each scenario (no difference across any tiers). Omnibus repeated-measures ANOVAs are reported as $F(df_1, df_2)$, where $df_1$ is the number of tiers minus one and $df_2$ reflects the degrees of freedom for within-subject error. F is the ratio of variance attributable to the actual differences across tiers relative to residual within-subject variance; partial eta-squared is an effect size, representing the proportion of that residual variance explained by differences across tiers. The subsequent tables show the McNemar's tests done to assess for significant change in proportion of treatment blue selections as well as the two-sample paired *t*-tests to detect statistically significant changes in confidence and perceived reliability. Specifically, across 12 patient scenarios, we conducted 58 planned within-subject paired comparisons (confidence and reliability ratings across tiers). To control for family-wise error rate, we applied a Holm-Bonferroni correction, yielding an adjusted significance threshold of $\alpha = 0.0009$. Results of the Shapiro-Wilks tests (done before conducting the 2-sample *t*-tests) for normality were all non-significant after adjusting for multiple hypothesis testing via the Bonferroni correction.

| **Scenario** | **Measure** | **Levels** | **Result** |
|---|---|---|---|
| **A** | Confidence | T1–T2–T3 | $F(2,62) = 4.33, p = 0.017, \eta^2p = 0.123$ † |
| | Reliability | T2–T3 only | 2 levels; ANOVA n/a |
| **B** | Confidence | T1–T2–T3 | $F(2,62) = 7.16, p = 0.002, \eta^2p = 0.188$ †† |
| | Reliability | T2–T3 only | 2 levels; ANOVA n/a |
| **C** | Confidence | T1–T2–T3 | $F(2,62) = 7.21, p = 0.002, \eta^2p = 0.189$ †† |
| | | T1–T2–T3–T4C | $F(3,93) = 4.62, p = 0.005, \eta^2p = 0.130$ †† |
| | Reliability | T2–T3–T4C | $F(2,62) = 18.63, p < .001, \eta^2p = 0.375$ ††† |
| **D** | Confidence | T1–T2–T3 | $F(2,62) = 3.54, p = 0.035, \eta^2p = 0.102$ † |
| | Reliability | T2–T3 only | 2 levels; ANOVA n/a |
| **E** | Confidence | T1–T2–T3 | $F(2,62) = 17.25, p < .001, \eta^2p = 0.358$ ††† |
| | | T1–T2–T3–T4C | $F(3,93) = 20.42, p < .001, \eta^2p = 0.397$ ††† |
| | | T1–T2–T3–T4NC | $F(3,93) = 11.25, p < .001, \eta^2p = 0.266$ ††† |
| | Reliability | T2–T3–T4C | $F(2,62) = 31.05, p < .001, \eta^2p = 0.500$ ††† |
| | | T2–T3–T4NC | $F(2,62) = 23.47, p < .001, \eta^2p = 0.431$ ††† |
| **F** | Confidence | T1–T2–T3 | $F(2,62) = 3.41, p = 0.039, \eta^2p = 0.099$ † |
| | Reliability | T2–T3 only | 2 levels; ANOVA n/a |
| **G** | Confidence | T1–T2–T3 | $F(2,62) = 1.16, p = 0.319, \eta^2p = 0.036$ ns |
| | Reliability | T2–T3 only | 2 levels; ANOVA n/a |
| **H** | Confidence | T1–T2–T3 | $F(2,62) = 20.24, p < .001, \eta^2p = 0.395$ ††† |
| | Reliability | T2–T3 only | 2 levels; ANOVA n/a |
| **I** | Confidence | T1–T2–T3 | $F(2,62) = 4.97, p = 0.010, \eta^2p = 0.138$ †† |
| | | T1–T2–T3–T4C | $F(3,93) = 3.42, p = 0.020, \eta^2p = 0.099$ † |
| | | T1–T2–T3–T4NC | $F(3,93) = 10.16, p < .001, \eta^2p = 0.247$ ††† |
| | Reliability | T2–T3–T4C | $F(2,62) = 17.17, p < .001, \eta^2p = 0.356$ ††† |
| | | T2–T3–T4NC | $F(2,62) = 28.04, p < .001, \eta^2p = 0.475$ ††† |
| **J** | Confidence | T1–T2–T3 | $F(2,62) = 18.01, p < .001, \eta^2p = 0.367$ ††† |
| | Reliability | T2–T3 only | 2 levels; ANOVA n/a |
| **K** | Confidence | T1–T2–T3 | $F(2,62) = 2.54, p = 0.087, \eta^2p = 0.076$ ns |
| | Reliability | T2–T3 only | 2 levels; ANOVA n/a |
| **L** | Confidence | T1–T2–T3 | $F(2,62) = 9.14, p < .001, \eta^2p = 0.228$ ††† |
| | Reliability | T2–T3 only | 2 levels; ANOVA n/a |

† $p < .05$ †† $p < .01$ ††† $p < .001$ ns = not significant. $\eta^2p$ benchmarks: small ≥ .01, medium ≥ .06, large ≥ .14. TX refers to Tier X

*Appendix Table 3*: Omnibus Repeated-Measures ANOVA by Scenario. For scenarios A–L, a one-way repeated-measures ANOVA tested whether confidence differed across tiers. Reliability ANOVAs are reported for scenarios C, E, and I (Tier 4 data available). Scenarios with only two reliability levels (Tier 2–Tier 3) are not amenable to ANOVA and are omitted.

| Measure | Tier 1 | Tier 2 | Tier 3 |
|---|---|---|---|
| **Scenario A** | | | |
| **Red (n)** | 32 | 32 | 30 |
| *McNemar T1→T2: $p = 1.000$ [0 switched (0.0%)]* | | | |
| *McNemar T2→T3: $p = 0.480$ [2 switched (6.2%)]* | | | |
| **Blue (n)** | 0 | 0 | 2 |
| **Confidence M (SD)** | 7.34 (1.21) | 7.53 (1.32) | 7.84 (1.19) |
| *T1→T2: $t(31) = -1.18$, $p = 0.245$, $d = -0.21$* | | | |
| *T2→T3: $t(31) = -1.97$, $p = 0.057$, $d = -0.35$* | | | |
| *T1→T3: $t(31) = -2.55$, $p = 0.016$, $d = -0.45$** | | | |
| **Reliability M (SD)** | — | 7.25 (1.32) | 7.44 (1.46) |
| *T2→T3: $t(31) = -0.83$, $p = 0.414$, $d = -0.15$* | | | |

| Measure | Tier 1 | Tier 2 | Tier 3 |
|---|---|---|---|
| **Scenario B** | | | |
| **Red (n)** | 30 | 23 | 24 |
| *McNemar T1→T2: $p = 0.046$ [9 switched (28.1%)]** | | | |
| *McNemar T2→T3: $p = 1.000$ [1 switched (3.1%)]* | | | |
| **Blue (n)** | 2 | 9 | 8 |
| **Confidence M (SD)** | 7.12 (2.09) | 6.03 (1.98) | 6.59 (1.92) |
| *T1→T2: $t(31) = 3.40$, $p = 0.002$, $d = 0.60$** | | | |
| *T2→T3: $t(31) = -4.76$, $p < .001$, $d = -0.84$**** | | | |
| *T1→T3: $t(31) = 1.46$, $p = 0.155$, $d = 0.26$* | | | |
| **Reliability M (SD)** | — | 5.88 (1.62) | 6.69 (1.65) |
| *T2→T3: $t(31) = -4.33$, $p < .001$, $d = -0.77$**** | | | |

| Measure | Tier 1 | Tier 2 | Tier 3 | Tier 4C |
|---|---|---|---|---|
| **Scenario C** | | | | |
| **Red (n)** | 32 | 11 | 10 | 9 |
| *McNemar T1→T2: $p < .001$ [21 switched (65.6%)]**** | | | | |
| *McNemar T2→T3: $p = 1.000$ [1 switched (3.1%)]* | | | | |
| **Blue (n)** | 0 | 21 | 22 | 23 |
| **Confidence M (SD)** | 7.22 (1.72) | 6.28 (1.57) | 6.78 (1.74) | 7.00 (1.90) |
| *T1→T2: $t(31) = 3.35$, $p = 0.002$, $d = 0.59$** | | | | |
| *T2→T3: $t(31) = -2.70$, $p = 0.011$, $d = -0.48$** | | | | |
| *T1→T3: $t(31) = 1.65$, $p = 0.109$, $d = 0.29$* | | | | |
| *T3→T4C: $t(31) = -1.23$, $p = 0.229$, $d = -0.22$* | | | | |
| **Reliability M (SD)** | — | 5.88 (1.48) | 6.59 (1.58) | 7.09 (1.84) |
| *T2→T3: $t(31) = -4.40$, $p < .001$, $d = -0.78$**** | | | | |
| *T3→T4C: $t(31) = -2.55$, $p = 0.016$, $d = -0.45$** | | | | |

| Measure | Tier 1 | Tier 2 | Tier 3 |
|---|---|---|---|
| **Scenario D** | | | |
| **Red (n)** | 32 | 4 | 3 |
| *McNemar T1→T2: $p < .001$ [28 switched (87.5%)]**** | | | |
| *McNemar T2→T3: $p = 1.000$ [3 switched (9.4%)]* | | | |
| **Blue (n)** | 0 | 28 | 29 |
| **Confidence M (SD)** | 7.28 (1.59) | 6.28 (1.73) | 7.03 (1.89) |
| *T1→T2: $t(31) = 2.24$, $p = 0.033$, $d = 0.40$** | | | |
| *T2→T3: $t(31) = -3.21$, $p = 0.003$, $d = -0.57$** | | | |
| *T1→T3: $t(31) = 0.55$, $p = 0.585$, $d = 0.10$* | | | |
| **Reliability M (SD)** | — | 6.03 (1.53) | 6.59 (1.78) |
| *T2→T3: $t(31) = -2.15$, $p = 0.039$, $d = -0.38$** | | | |

* $p < .05$ (uncorrected) *** $p < .05$ (Bonferroni-corrected; paired t-tests: $\alpha = 0.0009$ over 54 comparisons; McNemar's tests: $\alpha = 0.0021$ over 24 comparisons). $N = 32$. Paired t-tests for continuous measures; McNemar's tests for treatment switching.

*Appendix Table 4*: Within-patient paired analysis: treatment selection results and associated confidence and reliability metrics for scenarios A-D. In these scenarios, the patient meets the inclusion criteria for the RCT, and the ML model is trained on similar patients.

| Measure | Tier 1 | Tier 2 | Tier 3 | Tier 4C | Tier 4NC |
|---|---|---|---|---|---|
| **Scenario E** | | | | | |
| **Red (n)** | 27 | 28 | 27 | 27 | 26 |
| McNemar T1→T2: $p = 1.000$ [1 switched (3.1%)] | | | | | |
| McNemar T2→T3: $p = 1.000$ [1 switched (3.1%)] | | | | | |
| **Blue (n)** | 5 | 4 | 5 | 5 | 6 |
| **Confidence M (SD)** | 5.97 (2.13) | 6.50 (1.98) | 7.16 (1.76) | 7.56 (1.85) | 5.81 (1.77) |
| T1→T2: $t(31) = -2.79, p = 0.009, d = -0.49$* | | | | | |
| T2→T3: $t(31) = -4.12, p < .001, d = -0.73$*** | | | | | |
| T1→T3: $t(31) = -4.79, p < .001, d = -0.85$*** | | | | | |
| T3→T4C: $t(31) = -3.04, p = 0.005, d = -0.54$* | | | | | |
| T3→T4NC: $t(31) = 5.16, p < .001, d = 0.91$*** | | | | | |
| **Reliability M (SD)** | — | 6.00 (1.59) | 6.78 (1.66) | 7.28 (1.75) | 5.06 (1.74) |
| T2→T3: $t(31) = -4.88, p < .001, d = -0.86$*** | | | | | |
| T3→T4C: $t(31) = -3.71, p < .001, d = -0.66$*** | | | | | |
| T3→T4NC: $t(31) = 6.03, p < .001, d = 1.07$*** | | | | | |

| Measure | Tier 1 | Tier 2 | Tier 3 |
|---|---|---|---|
| **Scenario F** | | | |
| **Red (n)** | 25 | 15 | 14 |
| McNemar T1→T2: $p = 0.009$ [12 switched (37.5%)]* | | | |
| McNemar T2→T3: $p = 1.000$ [3 switched (9.4%)] | | | |
| **Blue (n)** | 7 | 17 | 18 |
| **Confidence M (SD)** | 6.34 (2.04) | 5.66 (1.93) | 6.28 (1.67) |
| T1→T2: $t(31) = 2.20, p = 0.035, d = 0.39$* | | | |
| T2→T3: $t(31) = -3.90, p < .001, d = -0.69$*** | | | |
| T1→T3: $t(31) = 0.17, p = 0.864, d = 0.03$ | | | |
| **Reliability M (SD)** | — | 5.75 (1.65) | 6.50 (1.61) |
| T2→T3: $t(31) = -3.34, p = 0.002, d = -0.59$* | | | |

| Measure | Tier 1 | Tier 2 | Tier 3 |
|---|---|---|---|
| **Scenario G** | | | |
| **Red (n)** | 25 | 7 | 5 |
| McNemar T1→T2: $p < .001$ [18 switched (56.2%)]*** | | | |
| McNemar T2→T3: $p = 0.480$ [2 switched (6.2%)] | | | |
| **Blue (n)** | 7 | 25 | 27 |
| **Confidence M (SD)** | 6.12 (1.77) | 5.81 (1.87) | 6.38 (1.76) |
| T1→T2: $t(31) = 0.72, p = 0.475, d = 0.13$ | | | |
| T2→T3: $t(31) = -2.67, p = 0.012, d = -0.47$* | | | |
| T1→T3: $t(31) = -0.59, p = 0.559, d = -0.10$ | | | |
| **Reliability M (SD)** | — | 5.66 (1.47) | 6.53 (1.88) |
| T2→T3: $t(31) = -4.28, p < .001, d = -0.76$*** | | | |

| Measure | Tier 1 | Tier 2 | Tier 3 |
|---|---|---|---|
| **Scenario H** | | | |
| **Red (n)** | 26 | 3 | 1 |
| McNemar T1→T2: $p < .001$ [23 switched (71.9%)]*** | | | |
| McNemar T2→T3: $p = 0.480$ [2 switched (6.2%)] | | | |
| **Blue (n)** | 6 | 29 | 31 |
| **Confidence M (SD)** | 5.75 (1.88) | 6.38 (1.79) | 7.56 (1.16) |
| T1→T2: $t(31) = -2.06, p = 0.048, d = -0.36$* | | | |
| T2→T3: $t(31) = -4.72, p < .001, d = -0.83$*** | | | |
| T1→T3: $t(31) = -5.86, p < .001, d = -1.04$*** | | | |
| **Reliability M (SD)** | — | 5.91 (1.77) | 6.97 (1.49) |
| T2→T3: $t(31) = -6.34, p < .001, d = -1.12$*** | | | |

* $p < .05$ (uncorrected) *** $p < .05$ (Bonferroni-corrected; paired t-tests: $\alpha = 0.0009$ over 54 comparisons; McNemar's tests: $\alpha = 0.0021$ over 24 comparisons). N = 32. Paired t-tests for continuous measures; McNemar's test for treatment switching.

*Appendix Table 5*: Within-patient paired analysis: treatment selection results and associated confidence and reliability metrics for scenarios E-H. In these scenarios, the patient does not meet the inclusion criteria for the RCT, though the ML model is trained on similar patients.

| Measure | Tier 1 | Tier 2 | Tier 3 | Tier 4C | Tier 4NC |
| --- | --- | --- | --- | --- | --- |
| **Scenario I** | | | | | |
| **Red (n)** | 27 | 29 | 24 | 25 | 21 |
| McNemar T1→T2: $p = 0.480$ [2 switched (6.2%)] | | | | | |
| McNemar T2→T3: $p = 0.131$ [7 switched (21.9%)] | | | | | |
| **Blue (n)** | 5 | 3 | 8 | 7 | 11 |
| **Confidence M (SD)** | 5.84 (2.00) | 6.59 (1.90) | 5.88 (1.93) | 5.97 (2.04) | 4.94 (1.97) |
| T1→T2: $t(31) = -3.28$, $p = 0.003$, $d = -0.58$* | | | | | |
| T2→T3: $t(31) = 2.70$, $p = 0.011$, $d = 0.48$* | | | | | |
| T1→T3: $t(31) = -0.10$, $p = 0.919$, $d = -0.02$ | | | | | |
| T3→T4C: $t(31) = -0.39$, $p = 0.703$, $d = -0.07$ | | | | | |
| T3→T4NC: $t(31) = 3.48$, $p = 0.002$, $d = 0.62$* | | | | | |
| **Reliability M (SD)** | — | 5.97 (1.89) | 4.41 (2.21) | 4.94 (2.31) | 3.91 (2.12) |
| T2→T3: $t(31) = 5.31$, $p < .001$, $d = 0.94$*** | | | | | |
| T3→T4C: $t(31) = -2.79$, $p = 0.009$, $d = -0.49$* | | | | | |
| T3→T4NC: $t(31) = 1.97$, $p = 0.058$, $d = 0.35$ | | | | | |

| Measure | Tier 1 | Tier 2 | Tier 3 |
| --- | --- | --- | --- |
| **Scenario J** | | | |
| **Red (n)** | 21 | 28 | 22 |
| McNemar T1→T2: $p = 0.023$ [7 switched (21.9%)]* | | | |
| McNemar T2→T3: $p = 0.041$ [6 switched (18.8%)]* | | | |
| **Blue (n)** | 11 | 4 | 10 |
| **Confidence M (SD)** | 6.16 (2.07) | 7.28 (2.10) | 5.50 (2.00) |
| T1→T2: $t(31) = -4.67$, $p < .001$, $d = -0.83$*** | | | |
| T2→T3: $t(31) = 5.41$, $p < .001$, $d = 0.96$*** | | | |
| T1→T3: $t(31) = 2.03$, $p = 0.051$, $d = 0.36$ | | | |
| **Reliability M (SD)** | — | 5.97 (2.15) | 4.47 (1.92) |
| T2→T3: $t(31) = 5.15$, $p < .001$, $d = 0.91$*** | | | |

| Measure | Tier 1 | Tier 2 | Tier 3 |
| --- | --- | --- | --- |
| **Scenario K** | | | |
| **Red (n)** | 25 | 2 | 3 |
| McNemar T1→T2: $p < .001$ [23 switched (71.9%)]*** | | | |
| McNemar T2→T3: $p = 1.000$ [3 switched (9.4%)] | | | |
| **Blue (n)** | 7 | 30 | 29 |
| **Confidence M (SD)** | 5.88 (2.01) | 6.38 (1.56) | 5.53 (2.08) |
| T1→T2: $t(31) = -1.26$, $p = 0.217$, $d = -0.22$ | | | |
| T2→T3: $t(31) = 3.48$, $p = 0.002$, $d = 0.62$* | | | |
| T1→T3: $t(31) = 0.75$, $p = 0.458$, $d = 0.13$ | | | |
| **Reliability M (SD)** | — | 5.59 (1.70) | 4.75 (2.16) |
| T2→T3: $t(31) = 2.44$, $p = 0.020$, $d = 0.43$* | | | |

| Measure | Tier 1 | Tier 2 | Tier 3 |
| --- | --- | --- | --- |
| **Scenario L** | | | |
| **Red (n)** | 21 | 27 | 18 |
| McNemar T1→T2: $p = 0.114$ [10 switched (31.2%)] | | | |
| McNemar T2→T3: $p = 0.008$ [9 switched (28.1%)]* | | | |
| **Blue (n)** | 11 | 5 | 14 |
| **Confidence M (SD)** | 5.88 (2.12) | 6.78 (2.00) | 5.50 (1.78) |
| T1→T2: $t(31) = -3.72$, $p < .001$, $d = -0.66$*** | | | |
| T2→T3: $t(31) = 4.03$, $p < .001$, $d = 0.71$*** | | | |
| T1→T3: $t(31) = 1.06$, $p = 0.296$, $d = 0.19$ | | | |
| **Reliability M (SD)** | — | 5.50 (1.90) | 3.94 (1.97) |
| T2→T3: $t(31) = 5.08$, $p < .001$, $d = 0.90$*** | | | |

* $p < .05$ (uncorrected) *** $p < .05$ (Bonferroni-corrected; paired t-tests: $\alpha = 0.0009$ over 54 comparisons; McNemar's tests: $\alpha = 0.0021$ over 24 comparisons). N = 32. Paired t-tests for continuous measures; McNemar's test for treatment switching.

*Appendix Table 6*: Within-patient paired analysis: treatment selection results and associated confidence and reliability metrics for scenarios I-L. In these scenarios, the patient neither meets the inclusion criteria for the RCT nor is the ML model trained on similar patients.

**Appendix File 7: Exit interview and "think aloud" session of scenario K results**

| **Themes** | **Example** |
|---|---|
| **Variability in using RCT data** | |
| Providers evaluate different aspects of clinical data when using an RCT to make a treatment decision | "I worry about performance status" (P1)<br><br>"Looking at this drug, will there be an issue when the patient has CKD?" (P3)<br><br>"ECOG 3 is likely related to their disease, so I want to be aggressive with treatment" (P6) |
| There are different approaches to evaluating an RCT where your patient does not meet inclusion criteria | "It has to be a risk benefit discussion [with your patient]" (P2)<br><br>"You have to think, what are the patient preferences?" (P3)<br><br>"We routinely use treatments for patients who do not meet inclusion criteria, so I don't worry about that so much" (P6) |
| **Baseline trust in ML results** | |
| There is trust in ML data when it is presented without further information about how the model was trained | "I would have some trust in the data" (P1)<br><br>"I'm assuming the model is taking something specific about the patient into account…" (P5)<br><br>"I take it at face value, assume it's a perfect model, my confidence is high" (P6) |
| **Variability in interpretation of ML model quality** | |
| There is variability in how providers perceive the information that the ML model was not trained on similar patients | "This knocked down my confidence… There's no good, right answer<br><br>"I thought of [information that the ML model was not trained on similar patients] as a disclaimer and not a big deal since it had good performance" (P4)<br><br>"Even though the training dataset wasn't representative, it still performed well" (P5) |
| **Many variables affect real-life decision-making, and this study is good practice** | |
| Different levels of training may impact how providers integrate data | "People pay attention to, hey, this model performed really well… people don't really think about ML or the specific issues with ML. People in medicine are not trained to do that" (P3)<br><br>"I wonder if a Hematology Oncology fellow would choose differently; people tend to trust their own expertise" (P5)<br><br>"There's a difference in risk tolerance. I bet the Internal Medicine doctors in your study saw neutropenia and were worried, but the hematologists said, 'no problem'" (P6) |
| Participants identified this study as an important thought experiment | "[This study] makes me think about what I would want to do; how do I manage the data I'm given?" (P1)<br><br>"With Epic integrating ChatGPT there are going to be lots of times where we have to decide if a ML recommendation is good or not, if models represent our patients or not" (P2) |
| | "[This study] was fun, a good thought exercise to think through data" (P4) |

ML = machine learning, RCT = randomized controlled trial

Participant with their treatment selection with (confidence in selection / perceived reliability of ML model) on a Likert scale of 1-10 for each tier of information provided. Representative quotes are included.

| **Study Participant** | **Tier 1 RCT data only** | **Tier 2 ML data added** | **Tier 3 ML context added** |
|---|---|---|---|
| Fellow, Male (P1) | Red (4/NA) | Blue (5/7) | Red (4/8) |
| Resident, Male (P2) | Red (6/NA) | Blue (7/6) | Red (5/3) |
| Resident, Male (P3) | Red (6/NA) | Blue (3/5) | Red (5/3) |
| Fellow, Male (P4) | Blue (7/NA) | Blue (9/9) | Blue (8/9) |
| Resident, Female (P5) | Red (3/NA) | Blue (6/5) | Blue (6/7) |
| Attending, Male (P6) | Red (6/NA) | Blue (8/8) | Red (6/4) |

NA = not applicable; RCT = randomized controlled trial; ML = machine learning

**Appendix File 8: COREQ Checklist for exit interview and single scenario analysis**

| Topic | Item | Question | Answer |
|---|---|---|---|
| **Domain 1: Research team and reflexivity** | | | |
| Interviewer/facilitator | 1 | Which author/s conducted the interview group? | ZH and BDL |
| Credentials | 2 | What were the researcher's credentials? | ZH was a PhD candidate and BDL has a medical degree |
| Occupation | 3 | What was their occupation at the time of the study? | ZH was a PhD candidate and BDL was a hematology oncology fellow |
| Gender | 4 | Was the researcher male or female? | ZH is male and BDL is female |
| Experience and training | 5 | What experience or training did the researcher have? | BDL is trained in qualitative methods and ZH was trained by BDL |
| Relationship established | 6 | Was a relationship established prior to study commencement? | BDL had worked with one of the participating clinicians in the past |
| Participant knowledge interviewer | 7 | What did the participants know about the researcher? | Participants who worked with BDL in the past may have known BDL was interested in oncology informatics research |
| Interviewer characteristics | 8 | What characteristics were reported about the interviewer/facilitator? | No information about the researchers were provided during the interviews |
| **Domain 2: Study design** | | | |
| Methodological orientatio Theory | 9 | What methodological orientation was stated to underpin the study? | Braun and Clarke's methods for thematic analysis |
| Sampling | 10 | How were participants selected? | Participants were recruited from the simulated interventional study |
| Method of approach | 11 | How were the participants approached? | Participants were recruited via e-mail |
| Sample size | 12 | How many participants were in the study? | 6 participants |
| Non-participation | 13 | How many people refused to participate or dropped out? | 26 of 32 study participants (81.3%) did not respond to requests to participate in the follow-up interview |
| Setting of data collection | 14 | Where was the data collected? | Participants were allowed to participate remotely |
| Presence of non-participants | 15 | Was anyone else present besides the participants and researchers? | No |
| Description of sample | 16 | What are the important characteristics of the sample? | Key demographics are provided in the results section |
| Interview guide | 17 | Were questions, prompts, guides provided by the authors? Was it pilot tested? | The interview guide is provided in Supplementary File 5. It was not pilot tested. |
| Repeat interviews | 18 | Were repeat interviews carried out? | No |
| Audio/visual recording | 19 | Did the research use audio or visual recording to collect the data? | No, only field notes were recorded |
| Field notes | 20 | Were field notes made during and/of after the interview or focus group? | Yes, both researchers took field notes |
| Duration | 21 | What was the duration of the interviews group? | All interviews lasted between 15 and 30 minutes |
| Data saturation | 22 | Was data saturation discussed? | We were unable to reach data saturation |

| Transcripts returned | 23 | Were transcripts returned to participants for comment and/or correction? | No |
|---|---|---|---|
| **Domain 3: Analysis and findings** | | | |
| Number of data coders | 24 | How many data coders coded the data? | Two researchers coded the data |
| Description of the coding tre | 25 | Did authors provide a description of the coding tree? | All codes were reported as themes in the results section given the small sample size |
| Derivation of themes | 26 | Were themes identified in advance or derived from the data? | Potential themes were identified from the study results and used to create the interview guide. All final themes were derived directly from the data. |
| Software | 27 | What software, if applicable, was used to manage the data? | Not applicable |
| Participant checking | 28 | Did participants provide feedback on the findings? | No |
| Quotations presented | 29 | Were participant quotations presented to i the themes/findings? Was each quotation identified? | Yes and further details are available in Supplementary File 7 |
| Data and findings consistent | 30 | Was there consistency between the data presented and the findings? | Yes |
| Clarity of major themes | 31 | Were major themes clearly presented findings? | Yes, major themes are discussed in the results section of the manuscript |
| Clarity of minor themes | 32 | Is there a description of diverse cases or discussion of minor themes? | Yes, some minor themes are discussed in the results section of the manuscript |

Developed from: Tong A, Sainsbury P, Craig J. Consolidated criteria for reporting qualitative research (COREQ): a 32-item checklist for interviews and focus groups. *International Journal for Quality in Health Care*. 2007. Volume 19, Number 6: pp. 349-357.